\documentclass{aastex6}
 \usepackage{amsmath}
\usepackage{graphicx}
\usepackage{threeparttablex}
\newif\ifcolor
\begin{document}
\keywords{cosmology: large-scale structure of universe, galaxies: clusters: general}
\title{Structure in the 3D Galaxy Distribution:
III. Fourier Transforming the Universe: Phase and Power Spectra}
\author{Jeffrey D. Scargle, 
M. J. Way\altaffilmark{1,2}, P. R. Gazis\altaffilmark{3}}
\affil{NASA Ames Research Center, 
Astrobiology and Space Science Division,\\
Moffett Field, CA 94035, USA}
\email{Jeffrey.D.Scargle@nasa.gov, Michael.J.Way@nasa.gov, PGazis@sbcglobal.net}
\altaffiltext{1}{NASA Goddard Institute for Space Studies,
2880 Broadway, New York, NY, 10025, USA}
\altaffiltext{2}{Department of Astronomy and Space Physics,
Uppsala, Sweden}
\altaffiltext{3}{Thermo Fisher Scientific, San Jose, CA}

\shorttitle{Galaxy Distribution 3D Fourier Transform}


\begin{abstract}
We demonstrate the effectiveness 
of a relatively straightforward 
analysis of the complex 3D Fourier transform 
of galaxy coordinates derived from redshift surveys.
Numerical demonstrations of this approach 
are carried out on a volume-limited sample 
of the Sloan Digital Sky Survey redshift survey.
The direct unbinned transform yields 
a complex 3D data cube 
quite similar to that from 
the Fast Fourier Transform (FFT) 
of finely binned galaxy positions.
In both cases deconvolution of the 
sampling window function 
yields estimates of the true transform.
Simple power spectrum estimates 
from these transforms are 
roughly consistent with those using more 
elaborate methods.
The complex Fourier transform characterizes 
spatial distributional properties beyond the power spectrum 
in a manner different from (and we argue is
more easily interpreted than)
the conventional multi-point hierarchy.
We identify some 
threads of modern large scale inference methodology that will presumably 
yield detections in new wider and deeper surveys.
\end{abstract}
\clearpage

\section{Introduction: Perspective and Assumptions}\label{introduction}

This paper takes the background 
established by two previous publications
on the multiscale structure of the Universe
\citep[][Papers I and II, respectively]{WGS2011,WGS2015} 
in a different direction:
\emph{direct 3D Fourier analysis of the galaxy positions}.
The goal is to maximize three things: simplicity,
extraction of information from the data, 
and independence from assumptions and models.
The following list describes the principles
underlying this work.
These  items are largely methodological clarifications 
and not cosmological assumptions as such.
In many cases our approach 
differs  from previous work,
references to which are deferred to 
the next section.

\begin{enumerate}
\item {\bf A Limited Cosmological Sample.} 
Testing models  against finite-volume data
usually involves consideration of \emph{cosmic variance}.
To avoid the need to postulate properties of unobserved data
that is inherent in this notion, 
we here adopt a viewpoint nicely described
(but not necessarily endorsed) by \cite{peebles_1}:
\begin{quote}
``One can adopt the view that we have only one Universe, 
that we can see only part of it,
and that the analysis ought to be based on that part alone.''
\end{quote}
\noindent
See Section \ref{internal_variance}  
for an approach to cosmic variance using resampling techniques.

\item {\bf Nearly Noise-free Data.} 
For our purposes
the uncertainty due to observational errors in the SDSS data 
is essentially negligible (Section \ref{error_analysis}).
We thus do not follow the common practice of 
treating irregularities at small scales as noise (or as not 
``topologically persistent") with smoothing 
or other practices that destroy information.
Structure detected on all scales
carries significant information about the 
Universe.

\item {\bf Point Distribution, not a Continuous Field.}  
Galaxies are often assumed to trace  
some underlying continuous field -- e.g. 
an averaged luminous or dark matter density, 
or a probability density for the presence
of a galaxy. 
We address the
spatial distribution of galaxies without 
reference to any continuous field,
and treat galaxies as discrete entities whose
spatial distribution carries information
about the structure of the Universe.
This approach is consistent
with equal treatment of all 
galaxies --  i.e. massive galaxies are not
given more weight than light ones.

\item {\bf Summary Distributions. }  There are two 
qualitatively different approaches: 
\begin{itemize}
\item detailed representation and interpretation of local structures
\item estimation of a few summary, global distributional parameters
\end{itemize}
again nicely spelled out in \cite{peebles_1}.
In Papers I and II we opted for the former.
Here we
derive several Fourier analysis functionals 
with the goal of estimating 
important global parameters.

\item {\bf Gaussianity.}
We approach the search for signs of 
non-Gaussianity via the Fourier phase spectrum.
A complete characterization of Gaussianity is contained 
in the power spectrum. 
The information about non-Gaussianity contained
in the phase spectrum is organized
in a form that is relatively easy to interpret
(cf. Section \ref{non_gaussianity}).

\item {\bf Absolute Clustering.} 
Most previous analysis has treated spatial correlations 
as departures from the mean density
\citep[e.g.][]{yu_peebles,landy_szalay1993}.
In contrast density estimates here and in Papers I and II 
are absolute.
In fact as discussed in Section \ref{3D_direct}
subtraction of a reference value,
such as the mean, 
does not make sense for a direct Fourier transform 
as in Eq. (\ref{ft_a}).
In addition our approach avoids some technical 
problems \citep{jones}.\footnote{A related comment
applies to the standard way to estimate correlation
functions \citep{peebles_1}.
A formula  for the probability 
that a galaxy be found within volume element $dV$ 
at distance $r$ from a randomly chosen galaxy, 
$\delta P = n [ \ 1 + \xi(r) \ ] dV$,
with $n$ the mean density, 
is conventionally used to define 
the autocorrelation function $\xi(r)$ as
a measure of  clustering. This definition
yields a quantity explicitly in excess (or deficit) 
of the mean.
Its normalization, 
$\int  \xi(r)) dV = -{1 \over n}$, is negative due to the
hold-one-out procedure,
and is very small  
due to the referencing to the mean.
This can be awkward for fitting positive only models
of $\xi(r)$ (e.g. power laws) 
to cosmological data.}

\item {\bf Explicit Deconvolution of the 3D Selection Function.} 
Using standard Fourier transform methods we
avoid constructs such as  Monte Carlo
simulations of ``un-clustered'' points within the
selected volume. 
For example,
\cite{feldman_kaiser_peacock1994} state: 
``Our approach is to take the Fourier transform of the real
galaxies minus the transform of a synthetic catalog
with the same angular and radial selection function as 
the real galaxies but otherwise without structure.''
Perhaps this approach has meaning in the context
of a model based on underlying randomness, but
is not a prescription for deconvolving the selection function.
Further, in consonance with item 1 and 2, 
we avoid interpreting such variance as a measure of uncertainty.
While various deconvolution methods have been employed
for both CMB and galaxy data, we believe the direct
3D Fourier deconvolution described in 
Section \ref{window_deconvolution} is novel.

\item {\bf Use All the Data.} 
In order to maximize the information gleaned from
the analysis, where possible we use all of the data.
For example we do not discard galaxies near the edges
of the data space in order to simplify the shape of the window
function (Section \ref{3D_window}),
but the next item indicates one case where
we feel a cut on the data is justified.

\item {\bf Volume-Limited Samples.} 
Throughout, as in Papers I and II, we use well
defined volume-limited samples.
This minor violation of the previous item is made 
for the good reason that it avoids bias corrections  
necessary for a magnitude-limited sample.

\item {\bf Omitted Effects.} 
Due to the small radial depth 
of our relatively shallow volume limited sample 
(redshift $\le 0.12$) and our interest in the simplest 
analysis, we have neglected many processes 
known to affect the data, including evolution and nonlinear 
cosmological terms, peculiar velocities, 
gravitational lensing, and local and non local 
GR terms depending on Bardeen potentials
and their temporal derivatives \citep[][especially Fig. 1]{raccanelli2015}.
\end{enumerate}
\noindent
The few of these viewpoints that are nonstandard
are not meant as criticisms of other approaches.
The goal here is limited to investigating 
the simplest possible way to extract 
spatial frequency information,
largely avoiding model-specific assumptions
and concentrating more on the phase spectrum 
and less on the power spectrum.



The organization of the rest of this paper is as follows:
Following a brief survey of prior work
in Section \ref{previous_work}, 
Section \ref{3Dft}
provides explicit details of 
two different ways to compute Fourier transforms -- a
direct  unbinned approach
and a fast Fourier transform of galaxy coordinates
in 3D spatial bins -- and a simple procedure for
deconvolution of the sampling window from 
the estimated galaxy transform.
Section \ref{results}
gives examples of application of the 
complex 3D Fourier  transform, 
briefly dwelling on the
amplitude (power) spectrum 
but focusing on the phase spectrum 
as a measure of Gaussianity.
Section  \ref{error_analysis} 
briefly addresses uncertainties.
The epilog in Section \ref{epilog} 
provides a summary 
and pointers to three statistical techniques 
the should be useful in future research.
Two appendices 
provide a check of the analysis
and some MatLab code.

\section{Previous Work}
\label{previous_work}

A small part of the earlier relevant literature  can be found in 
\cite{L1953,G1965,yu_peebles,peebles_1,peebles_2,peebles_3}.
The cosmological importance of power spectrum analysis 
has recently been extensively discussed in \cite{cwz2015},
 especially in the context of galaxy redshift surveys 
 \cite[e.g.][Sec. 1.1]{vogeley_szalay_96}.
Fourier phases have been studied 
in connection with cosmic microwave background data
\citep[see e.g.][]{CNVM2003,NDV2003,NDV2004,CNC2004,NCON2005,CN2007,KSF2013,ferreira_magueijo1997}.
More recently phase analysis is beginning to be
applied to galaxy redshift data
\citep{HMSPSB2005,M2007,WBO2015,EBSN2015}.

Some relevant work on non-Gaussianity,
much in the context of CMB but with application to 
large scale -- or more appropriately multi-scale -- structure includes:
\cite{Hikage2006} who provide a general overview,
\cite{sefusatti2007} who discuss the bi-spectrum for high redshift galaxies,
\cite{Hikage2008} who discuss the application of Minkowski
functionals,
\cite{sanchez_coles2008} who estimate the power spectrum 
using Fourier methods based on work by 
\cite{feldman_kaiser_peacock1994},
\cite{martinez-gonzalez2009} and
\cite{lentati} for pulsar timing studies,
and also \cite{KSF2013}.
See \cite{coles2005} for application of Fourier methods 
to the 2dF galaxy redshift survey,
employing the Fourier based method of 
\cite{PVP2004}, a generalization of the
minimum variance method of 
\cite{feldman_kaiser_peacock1994}.  
\cite{coles2005}  derive power spectra and compare them 
to several empirical  \cite[e.g.][]{tegmark2004a} and theoretical results.
See \cite{kitaura2012} for derivation of some statistics
relevant to non-Gaussianity in galaxy clustering.
Other work on non-Gaussianity can be found in 
 \citep{CC2000,rocha2001,wcm,tegmark2004b}.
 

\cite{efstathiou_moody}
describe a method of recovering the three-dimensional power spectrum from measurements of the angular correlation function
applied to the APM galaxy survey -- 
one of the first large surveys using 
automatic plate measuring methods. 
See also \cite{querre} for discussion of the
galaxy distribution using multiscale methods
in general, and 3D implementations
of the  \'{a} trous algorithm, and the ridgelet 
and beamlet transforms in particular. 
Percival, Verde \& Peacock (2004, hereafter PVP) 
studied luminosity-dependent
galaxy clustering with 
spherically averaged Fourier analysis. 
\cite{coles2005} applied the Fourier based 
method of PVP to the  2dF galaxy redshift survey.
This approach in turn is 
a generalization of the
minimum variance method of 
\cite{feldman_kaiser_peacock1994}.
See recent papers 
\citep{slepian2015a,slepian2015b,slepian2015c,dore}
for estimation of three-point correlation functions
and their application to problems 
in dark matter cosmology.
\cite{alam2016} give a recent summary of relevant literature and
extensive analysis of data from the 
SDSS-III Baryon
Oscillation Spectroscopic Survey.

\clearpage

\section{3D Fourier Transforms}
\label{3Dft}
The sub-sections below describe the procedures  
used to compute the complex Fourier transform
of the galaxy distribution -- first reviewing the data
and then outlining direct and binned transforms of 
the galaxy positions and the corresponding data 
window, concluding with a Fourier-based 
procedure to deconvolve the effect of this window function.

\subsection{The Data}
\label{data}

Below we Fourier analyze the SDSS DR7 data
described in Papers I and II,
namely the NASA/Ames Research
Center SDSS Value Added Galaxy Catalog (AMES-VAGC).
Data Release
7 of the SDSS 
was augmented using the 
New York University
Value Added Catalog
(see Appendix A of Paper II
for details and references).
As a reminder, 
redshift $\zeta$, right ascension
$\alpha$ and declination $\delta$ 
were converted to rectangular 
Cartesian coordinates with the formulas
\begin{eqnarray}
x = \zeta cos( \delta ) cos( \alpha ) \nonumber \\
y = \zeta cos( \delta ) sin( \alpha ) \\
z = \zeta sin( \delta ) \nonumber
\end{eqnarray}
\noindent
and no cosmological corrections
were made in view of the low-redshift
nature of the sample.
The one small difference is that, 
since the analysis here does not involve
Voronoi tessellation, the omission of 
the small sample of galaxies near
the edges of the data space (Paper I,
Section 5.2.2) is unnecessary.
This slightly increases the sample size.
More significant is the resulting improved definition
of the edges of the data space,
of importance for the transform of the 
data window 
described in Section \ref{3D_window}.

\subsection{Fourier Transform of the Galaxy Distribution}
\label{3D_direct}

Let's start with the Fourier transform
of the data, keeping an eye toward 
preserving both directional and phase 
information.
The Fourier transform of any function $f( {\bf x} )$ 
defined over a 3D volume $V$
in ${\bf x} = ( x, y, z )$-space 
is, without specifying a normalization, 
\begin{equation}
F_{f}(\bf{k} ) = 
\int_{V} f( {\bf x} ) e^{ -i \bf{k} \cdot \bf{x} } d\bf{x}
\label{ft}
\end{equation}
\noindent
where $\bf{k}$ is the spatial frequency vector
$\bf{k} = (k_x, k_y, k_z)$ -- chosen so that
the linear scale (one full period of the sinusoid) 
corresponding to $k$ is $2 \pi / k$.

Following \cite{yu_peebles,peebles_1,peebles_2,peebles_3}
we account for the discreteness of the data  by
taking $f( {\bf x} )$ in Equation (\ref{ft}) 
as the sum of point locator functions, 
i.e. delta functions at the positions ${\bf x_{n}}$ of each of the galaxies:
\begin{equation}
f( \bf{x} ) = \sum_{n=1}^{N} \delta( \bf{x} - \bf{x}_n )  \ \ ,
\label{unit_sum}
\end{equation}
\noindent 
where the sum is over the $N$ galaxies included in the
volume limited sample.
(See also \cite{BBKS1986} for a similar representation
in terms of peaks -- local 3D maxima -- of density.)
The resulting galaxy Fourier transform is simply
\begin{equation}
F( \bf{k} ) = 
\int_{V} \sum_{n=1}^{N}  \ 
  \delta( \bf{x} - \bf{x}_n )  e^{ -i \bf{k} \cdot \bf{x} } d\bf{x} = 
  \sum_{n=1}^{N} \ 
    e^{ -i \bf{k} \cdot \bf{x_n} }  \ , 
 \label{ft_a}
  \end{equation}
   \noindent
where the dot is the vector scalar product.
In component notation
  \begin{equation}
F (k_x, k_y, k_z )    =    \sum_{n=1} ^{N}  \ e^{ -i ( k_x x_n + k_y y_n + k_z z_n )} \ .
\label{ft_b}
\end{equation}
\noindent
To examine the behavior of the transform for
small frequencies, 
after defining the galaxy centroid as 
\begin{equation}
< x > \equiv {1 \over N} \sum_{n=1}^{N} \bf{x}_n \ , 
\end{equation}
a simple computation gives 
\begin{equation}
\underset{\bf{k} \rightarrow 0 } { F( \bf{k} ) } 
\sim 
N [ 1 -i \bf{k} \ \cdot<\bf{x}> + {1 \over 2}   <( \bf{k} \cdot \bf{x_n}) ^ 2 >  + \dots ] 
 \label{fourier_transform_expand}
 \end{equation}
 \noindent
Thus the normalization is $F(0) = N$,
and one sees that the transform is smooth at $\bf{k}=0$.
With this form of the Fourier transform there is no analog
of subtracting the mean value to bring the power spectrum 
to zero at zero frequency.
The nearest thing to this procedure is to 
remove the linear term in 
Eqs. (\ref{fourier_transform_expand}) 
by referring the coordinates to the centroid;
but the higher order terms are still smooth at the origin.

\begin{figure}[htb]
\includegraphics[scale=1]{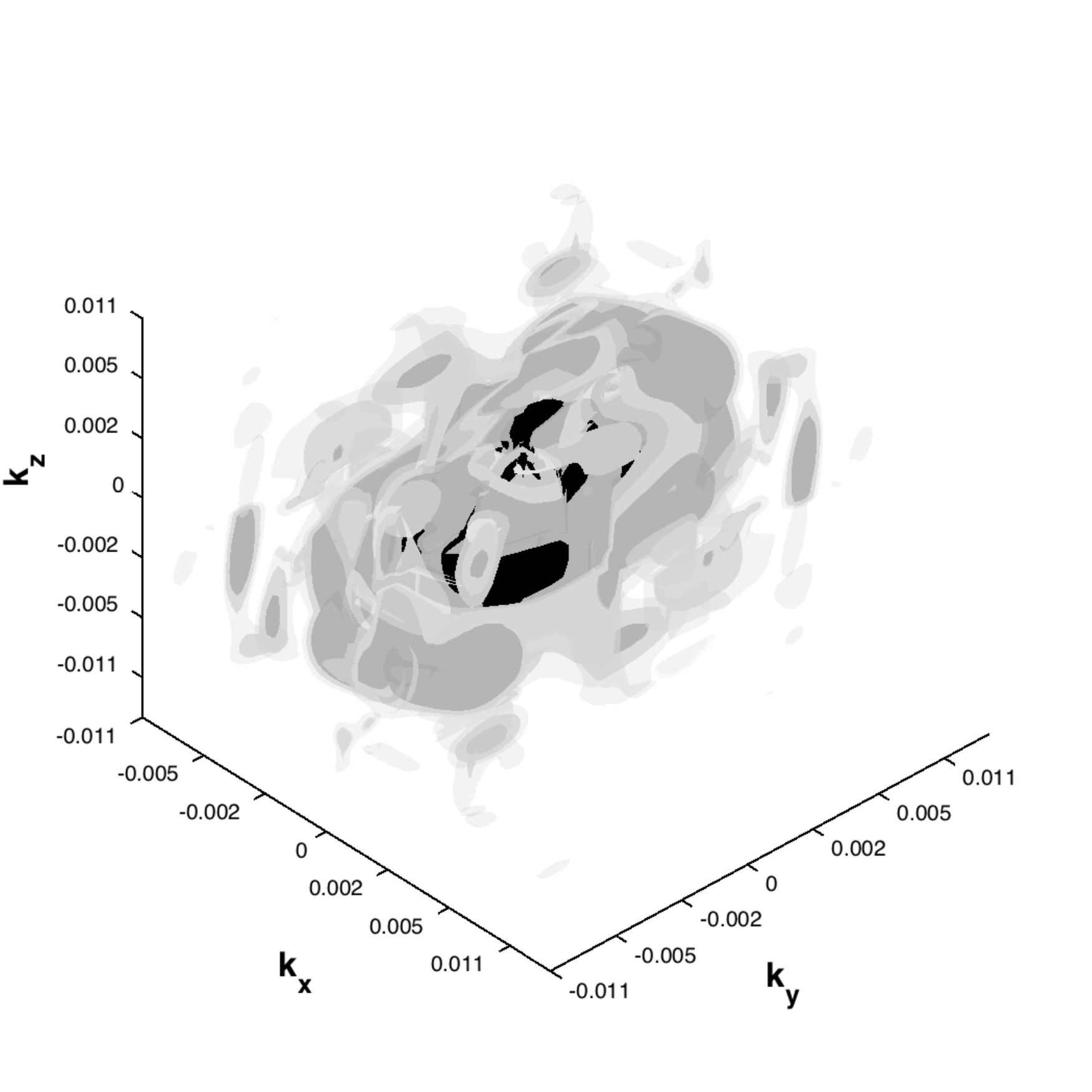}
\caption{Isosurface plot of
the Fourier power spectrum.
The x,y and z coordinates are 
proportional to the base-10 log of spatial frequency, 
but labeled with the value of $k$ in units of h/Mpc.
The powers, 
in order of decreasing opacity of the iso-surface and 
expressed as fractions of the zero-frequency power $N^{2}$
are $0.8357,   0.7612,   0.7450$ and $ 0.7288$.
These levels were chosen to make this display
informative.
}
\label{ft_universe}
\end{figure}

The formula in Equation
(\ref{ft_b}) 
is easily evaluated for any galaxy 
sample, in time of order $N \times N_{k}^{3}$,
i.e. the product of the number of galaxies 
and the total number of frequencies ($N_{k}$ is 
the number of frequencies in a single coordinate direction).
It treats all galaxies 
as identical points.
Through its response to 
crowding together of galaxies in various regions 
it is sensitive to the local number density of galaxies, 
but by choice not to mass density.
Figure \ref{ft_universe} 
displays the 3D structure  of 
the 
Fourier power spectrum
$P({\bf k}) = P (k_x, k_y, k_z ) = | F (k_x, k_y, k_z ) |^{2}$.
Since the definition in equation (\ref{ft_a})
does not include a factor $1/V$,
powers throughout are dimensionless
and normalized to $P(0,0,0)  = N^{2}$.
Conversion to physical units (per unit volume) is easily made
from the value $9.47 \times 10^{7} Mpc^{3}$
for the volume of the convex hull of the data sample.
The higher frequencies roughly speaking
form an isotropic but somewhat irregular 
shell around the inner core (the black
shape at the center of the plot) of
low-frequency or large-scale structure.
Spectral quantities derived using equation
(\ref{ft_b})
will be called \emph{direct}, as opposed to \emph{binned} for
those from the methods in Section \ref{3D_binned}.
Appendix A describes the use of the inverse Fourier transform 
to check how well Eq. (\ref{ft_b}) represents the raw data.

\clearpage

\subsection{Fourier Transform of the Data Window}
\label{3D_window}

The data from a survey of a given volume $V$ 
can be thought of as the product of the actual 
3D spatial distribution of 
galaxies multiplied by a 3D spatial window,
or selection function, given by:
\begin{eqnarray}
S(\bf{x}) = & 1  \ \mbox{for} \  {\bf x}  \in V \\
                 & \  \  \ 0 \ \  \mbox{otherwise}  \ . \nonumber
\label{fourier_window}
\end{eqnarray}
\noindent
This window can be defined by the 2D footprint of the survey on the sky
combined with the 1D redshift selection function.
Here we use the corresponding volume in terms of rectangular
coordinates $\bf{x,y,z}$.
This approach ignores 
the variation of the redshift selection 
over the relatively narrow redshift range of our data.
Of course any survey
has this and additional selections, 
not considered here. 

By the well known convolution theorem \citep{bracewell} the 
Fourier transformation of this product relation 
yields the fact that 
the Fourier transform of the survey data, 
Eq. (\ref{ft_a}), 
is the transform of the actual 
distribution convolved with the
Fourier window function, defined as the transform of the selection function: 
\begin{equation}
F_{\mbox{window}}(\bf{k} ) = 
\int_{-\infty}^{\infty}  S(\bf{x}) e^{ -i \bf{k} \cdot \bf{x} } d\bf{x} =
\int_{V}  e^{ -i \bf{k} \cdot \bf{x} } d\bf{x} \ \ .
\label{ft_window}
\end{equation}
\noindent
In order to compute this integral exactly 
one could discard some of the data
and redefine $V$ as a simplified subset of
the actual data space, such as a figure
with planar boundaries. 
Here we wish to compute $F_{\mbox{window}}(\bf{k} )$
where $V$ is the actual 3D data space of the redshift survey.
Note that the linearity of equation (\ref{ft})
means that the Fourier transform can be evaluated
as a sum of transforms over 
the elements of any \emph{partition} of $V$; i.e. 
for any $f$ 
\begin{equation}
F_{f}(\bf{k} ) = \sum _{n}
\int_{V_n} f( {\bf x} ) e^{ -i \bf{k} \cdot \bf{x} } d\bf{x} \ ,
\label{ft_linear}
\end{equation}
\noindent
where \{${V_n}, \hskip .1in n = 1, 2, \dots $\}
is a set of disjoint volumes the union of which
is the full observation space $V$.
It is convenient here to partition $V$
into a set of rectangular parallelepipeds,
or \emph{cuboids}.
The contribution of a cuboid $C$,
i.e. the volume defined by 
\begin{eqnarray}
  S(\bf{x}) & = 1  & \ \ \ x_a \le x \le x_b;
                                                  \ y_a \le y \le y_b;
                                                 \ z_a \le z \le z_b  \\
    & = 0 &   \ \ \ \mbox{otherwise}\nonumber   \ ,
\end{eqnarray}
\noindent
can be found exactly as a function of its
bounding $xyz$ coordinates $x_a, x_b$ etc.
The Fourier transform of such a cuboid is just
\begin{eqnarray}
F_{C}(\bf{k} ) & =  &
\int_{C}  e^{ -i \bf{k} \cdot \bf{x} } d\bf{x} \\
 &  = &
\int_{x_a}^{x_b} e^{ -i k_x x } dx 
\int_{y_a}^{y_b} e^{ -i k_y y } dy 
\int_{z_a}^{z_b} e^{ -i k_z z } dz \\
&  = &
({e^{ -i k_x x_b} - e^{ -i k_x x_a } \over -i k_x})
({e^{ -i k_y y_b} - e^{ -i k_y y_a } \over -i k_y})
({e^{ -i k_z z_b} - e^{ -i k_z z_a } \over -i k_z})
\label{ft_cuboid}
\end{eqnarray}
\noindent

Now let's approximate the data space 
with a refined partition as follows:
Construct a grid of equal squares covering the
projection of $V$ onto 
the x-y plane, with a fine spacing  
$\Delta = x_b - x_a = y_b - y_a$.
To define a cuboid it remains to specify 
$z_a$ and $z_b$.
We take these as the $z$-coordinates 
at which a vertical line
through the center of the square and 
parallel to the $z$-axis intersects 
the convex hull of the full set of galaxy positions.
It is easy to see that each such line intersects the
convex hull in either 2 or 0 facets; in the latter case
the cuboid is entirely outside the data set and is ignored.

\begin{figure}[htb]
\includegraphics[scale=1.2]{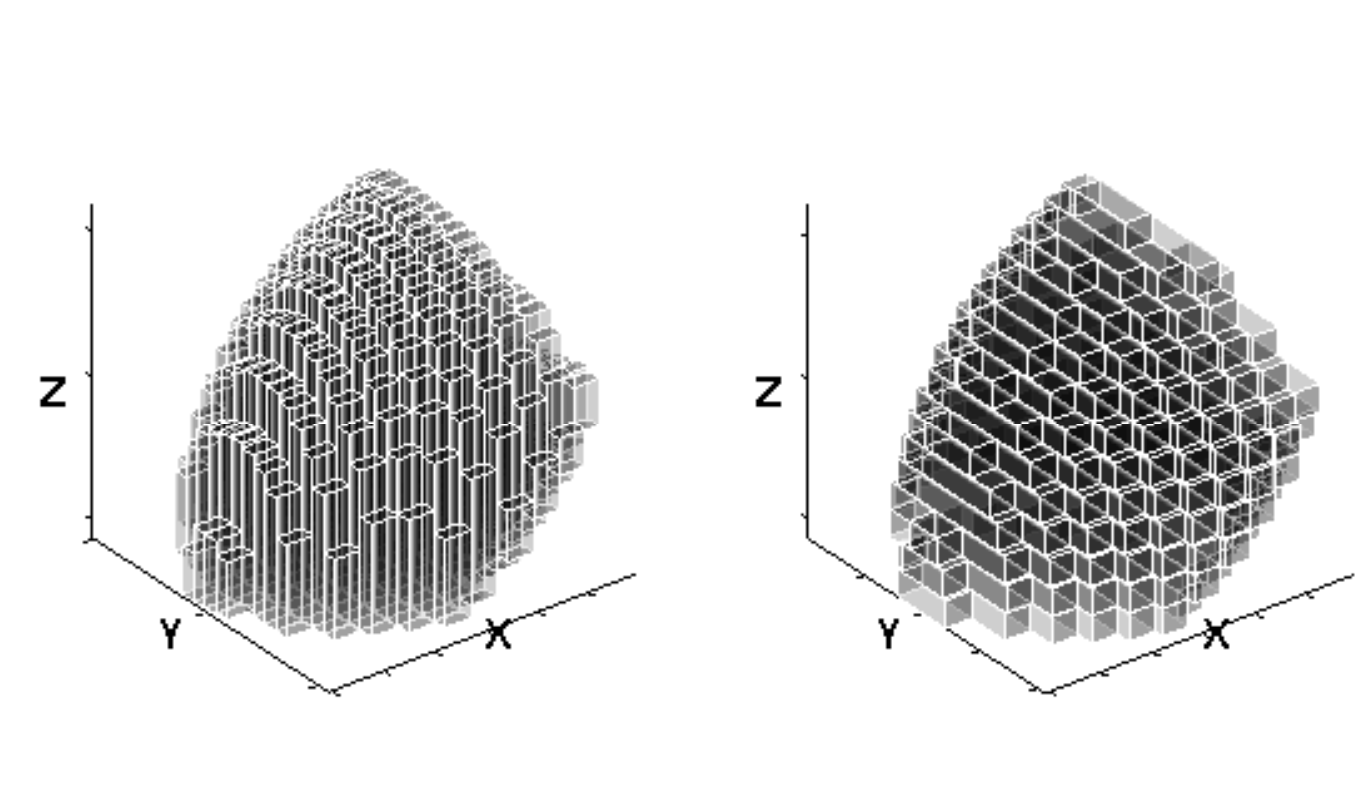}
\caption{Two sample partitions of the 
convex hull of the SDSS data into cuboids 
with transverse size .01 redshift units.
The long-axes of the cuboids are parallel
to the z-axis and y-axis. 
These crude partitions are for illustration only; 
those used in the analysis are much finer.
}
\label{cuboid_all}
\end{figure}

Figure \ref{cuboid_all} shows relatively crude partitions 
of the actual data space with the long axes of the cuboids
in two different directions.  If the transverse 
dimensions of the cuboids are sufficiently small 
the partition approaches an exact coverage of the
overall convex hull and the resulting window Fourier transform
is independent of the cuboid orientation.
For the grid size equal to .0001 redshift units (0.416 Mpc) 
the sum of the volumes of the cuboids matches the
exact volume of the convex hull to one part in $10^8$,
not surprising since this computation is equivalent to the
elementary integral calculus procedure for 
computing the volume
of the convex hull.
Putting all this together,
the result will be used to correct 
the galaxy Fourier transform 
for the effects of the data window -- cf. Section \ref{window_deconvolution}.
Appendix A describes a way to check how well 
the transform approximates the actual selection function,
and Appendix B describes MatLab code implementing
the Fourier transforms, including
the deconvolution of the window transform,
with a link to the code and data files in electronic form.

\subsection{{Fast Fourier Transform of the Binned Galaxy Distribution}}
\label{3D_binned}

Another way to estimate 
the Fourier transform
is to construct a 3D histogram of the
galaxy positions in a spatial grid of 3D bins, or
 \emph{voxels}.
To use the fast Fourier transform 
each voxel must be 
a cube with the same 
small size $\Delta x = \Delta y = \Delta z$.
This procedure discards some information,
due to rounding of galaxy coordinates and
placing some close pairs in the same voxel.
Table \ref{fft_statistics} summarizes 
statistics for three grid sizes.
Columns 5-8, giving
the maximum number of galaxies in any one 
voxel and the fractions with 0, 1, 
and 2 or more galaxies,
are useful in assessing the information
loss in this binning.
Ideally the fraction with more than one
galaxy would be zero, leaving coordinate 
truncation as the only error. 
Computation with 512 bins in each coordinate --
case (c), with 135,005,697 voxels -- 
seems to be the largest 
feasible with current personal computers.
As the accuracy of the computation 
increases more and more voxels are empty,
since the number not empty cannot 
exceed the number of galaxies.
The last column gives the fraction of voxels
empty because they are outside the
survey volume;
these values are large 
due to the shape of the survey and 
because we zero-padded it for 
good frequency resolution.
\begin{table}[htp]
\caption{Statistics for the Binned Fourier Transform}
\label{summary_structures_sdss}
\begin{center}
\begin{tabular}{   |  c |  c | r |  c |  c |  c  |  c | c | c |}
\hline 
Case &  $N_{\mbox{bins} }\footnote{Number of bins per dimension.}$ & $N_{\mbox{vox}}$\footnote{Number of voxels (millions).} & $\Delta$   \footnote{Linear dimension of voxels (Mpc).}  
& Max n \footnote{Maximum number of galaxies in a voxel.} 
& Fraction n=0 \footnote{Fraction of empty voxels.} 
& Fraction n=1 \footnote{Fraction of voxels containing one galaxy.} & Fraction n $>$ 1\footnote{Fraction of voxels containing more than one galaxy.}  & Fraction Outside\footnote{Fraction of voxels outside the convex hull of the data.} \\ 
\hline 
(a)  &  128 & 2.1  M & 6.8  & 32 &  0.9693 &.016519 &  .014221 &  0.8923 \\ 
(b)  &  256 & 16.8 M & 3.4  & 11 & 0.9938 & .004882 & .001300  & 0.9347 \\ 
(c)  &  512 & 134.2 M & 1.7 &  5 & 0.9991 &  .000869 &.000062 &  0.9347 \\ 
 \hline
\end{tabular}
\end{center}
\label{fft_statistics}
\end{table}
\clearpage

The Fourier transform of the window is simply
that of this bin array with unity
inside the sample volume and zero outside,
cf.  equation (\ref{fourier_window}).
Actually instead of the convex hull of the 
filled bins, for each dimension we assigned 
a unit value to each bin between  
the minimum and the maximum indices
of bins containing galaxies in all of the 
corresponding x-columns, y-columns and z-columns.
In practice this is essentially the same as the 
convex hull.
The inverse transform of the Fourier 
transform computed this
way is guaranteed to exactly reproduce the 
input counts-in-voxels, so there is no
point in numerically demonstrating the
accuracy of this representation as in
Appendix A for the direct transform.

\subsection{Deconvolution of the Data Window}
\label{window_deconvolution}

We approach correcting for the selection
function (or window) 
in a straightforward way.  
Functions of a 3D coordinate vector $\bf{x}$ 
related multiplicatively in the manner 
\begin{equation}
 q_{obs}({\bf x}) = q_{true} ({\bf x} )  q_{window}({\bf x} ) \ ,
\label{mult}
\end{equation}
\noindent
have spatial Fourier transforms related by
\begin{equation}
Q_{obs} ({\bf k} ) = Q_{true} ({\bf k} ) * Q_{window}({\bf k} ) \ , 
\label{convolution}
\end{equation}
\noindent
where $Q_{obs} (\bf{k} )$ is the Fourier transform of $q_{obs}(\bf{x})$, etc., and
$*$ means 3D convolution on the vector $\bf{k}$.
There are many \emph{deconvolution} 
techniques for finding $q_{true} (\bf{k} )$,
thus correcting for the window function,
but here the simple expedient of 
Fourier transforming
Eq. (\ref{convolution}) yields 
\begin{equation}
q_{true} (\bf{k} ) = F^{-1} {  F[ Q_{obs} (\bf{k} ) ] \over 
F[ Q_{window}(\bf{k} )  ]}
\label{deconvolution_eq1}
\end{equation}
\noindent
where $F$ and $F^{-1}$ are the forward and 
inverse Fourier transforms. 
In all numerical results presented here the 
MatLab (\copyright{MathWorks}) multidimensional 
functions \verb+fftn+ and \verb+ifftn+ were used 
for both the direct and binned cases.
This deconvolution method is sometimes avoided because
of worries about noise amplification and/or issues when
the denominator in eq. (\ref{deconvolution_eq1}) is zero 
(or small in absolute value), but here these issues do 
not cause any serious problems.

\section{Characterizing the Spatial Distribution of Galaxies}
\label{results}

We are now ready to use 
the above Fourier transform methods 
for global characterization
of the galaxy distribution.
It is useful to compare results
from the binned and unbinned
Fourier transforms.
Neither one is better in all aspects than the other.
Of course they both have limited spatial
frequency resolution, but their 
different data representations 
implement distinct approximations.
The binned approach
suffers from the information loss associated with
quantization of the galaxy coordinates.

\subsection{Fourier Power Spectrum}
\label{power}

\begin{figure}[htb]
\includegraphics[scale=1]{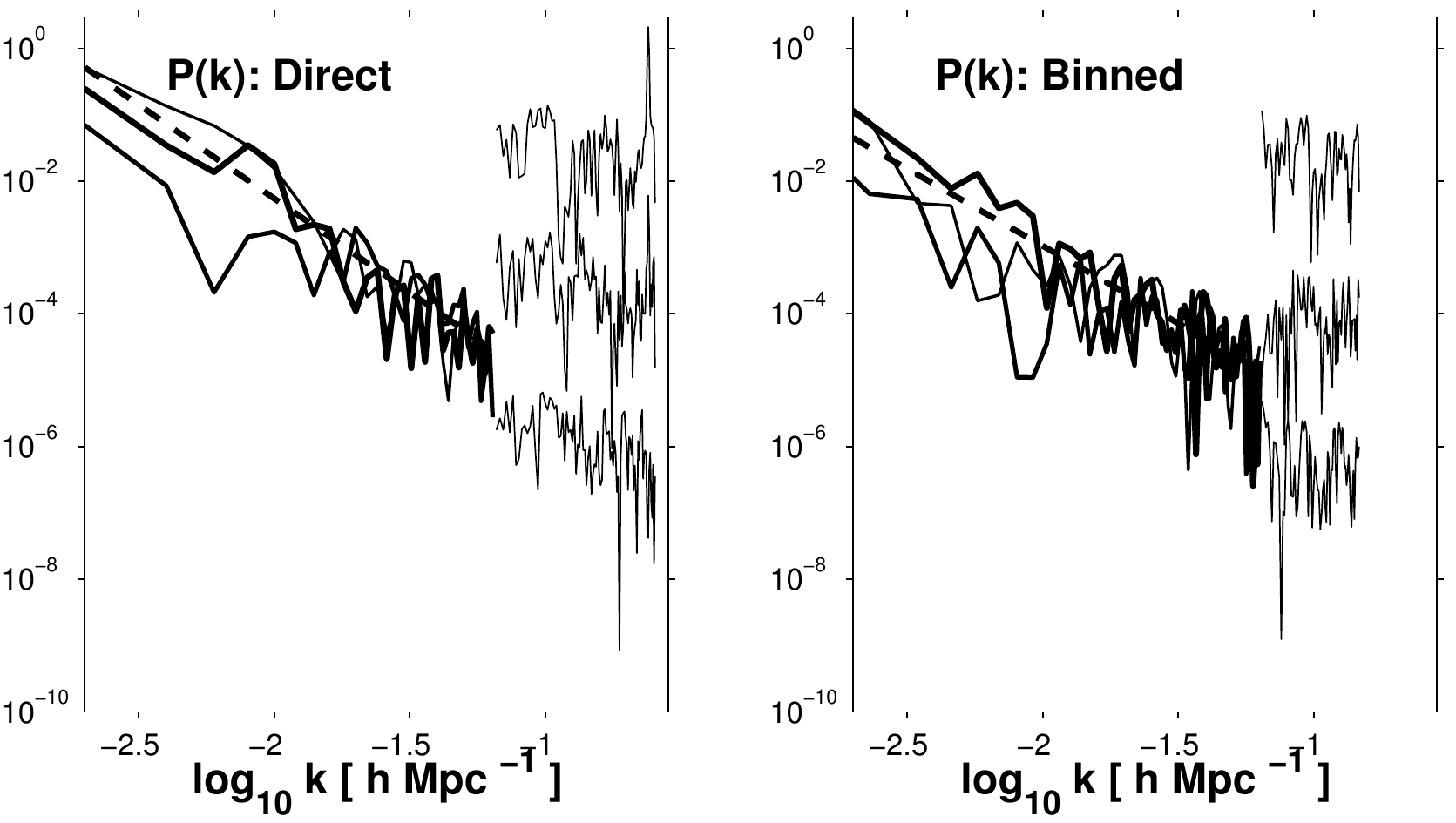}
\caption{Power Spectra from deconvolved 
direct (left) and binned (right) Fourier Transforms:
x, y and z powers in solid lines of increasing width.
As in Fig. \ref{ft_universe} the dimensionless power 
is shown divided by its zero-frequency value $N^{2}$ 
to yield P(0) = 1.
Above log $k$ of -1.2 
(spatial frequencies $> 0.063$) the powers
are multiplied by 3000, 10, and 0.1,
respectively, for clarity.
The dashed  straight  lines are power law
fits to the low frequency data
(averaged over the 3 directions) with
slopes -2.8 and -2.3 respectively.}
\label{fig_power_1}
\end{figure}

Figure \ref{fig_power_1}
shows the deconvolved 
power spectra for both methods:
direct as in Sections \ref{3D_direct} and \ref{3D_window}
and binned in Section \ref{3D_binned}.
The powers projected in three orthogonal directions,
$ | F( k_x, 0, 0 ) |^{2} $,
$ | F( 0, k_y, 0 ) |^{2} $, and 
$ | F( 0, 0, k_z ) |^{2} $,
are distinguished by lines of different widths.
These three power spectra 
share the same zero-frequency value,
namely $ | F( 0, 0, 0 ) |^{2} = N^{2}$,
and we have normalized the plotted
curves to unity at zero spatial frequency 
-- which of course is off-scale
on these log-log plots.
Comparison of the power spectra 
in different directions
provides a simple measure of isotropy.
The spectra at
lower spatial frequencies 
approximate the power-law 
dependence characteristic of 
red noise  \citep{MA1986}.
The straight (dashed) lines in this figure
are least-squares fits to the mean 
of the three power spectrum curves
in the interval below the cutoff
at log $k = -1.2$ 
mentioned in the caption;
the log-log slopes indicated
there
are not far from the 
common red noise value
of $\approx -2$.
The scatter reflecting 
high variability at small scales
motivates the vertical shifts
in the higher frequency part of these plots,
at the same cutoff used for the power-law fits.

\begin{figure}[htb]
\includegraphics[scale=1]{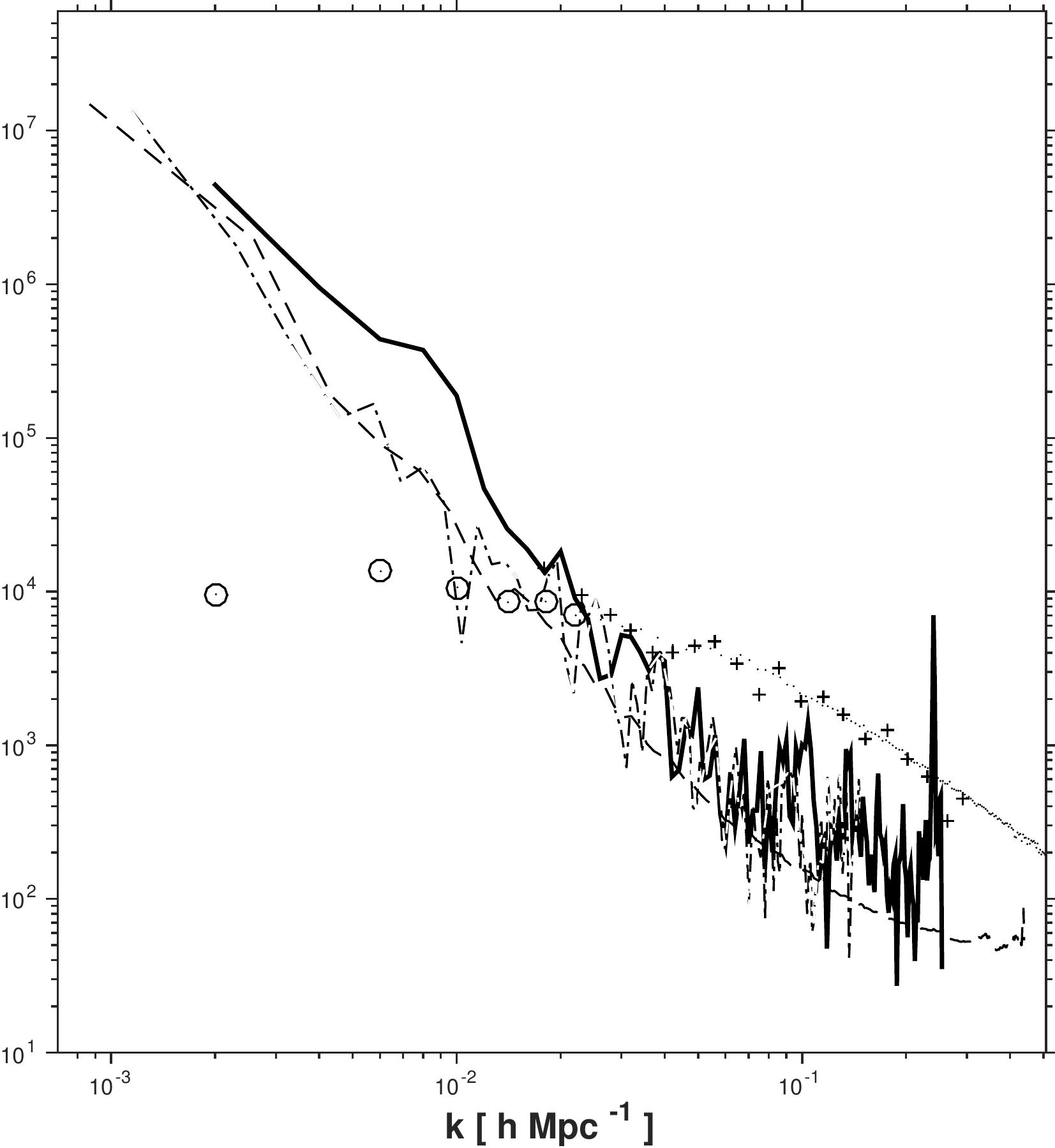}
\caption{Power Spectrum Comparison.
Solid line: power from average direct Fourier transform
(eq. \ref{ft_b}).
Dots-dashes: average binned FFT.
Dashes: average
of the direct powers at all of the 
frequencies falling in a
given 1D spherical volume in ${\bf k}$ space.
{\bf Our power spectra are renormalized to units of 
$(h^{-1} Mpc)^{3}$ for comparison with the other authors,
and }corrected for the
selection window (cf. Section \ref{window_deconvolution}). 
The spatial frequencies and powers from columns 1 and 2 of 
Table 2 in
\cite{tegmark2004b} are plotted 
as plus signs (+), and those of \cite{PNE2007} as small dots
(but with the lowest 6 frequencies emphasized 
by circumscribed circles).
}
\label{fig_power_2}
\end{figure}

Figure \ref{fig_power_2} plots our power spectra
against those from some other authors.
In interpreting the figure and 
assessing this comparison the reader should
bear in mind both the simplicity of our method -- 
using the unadorned Fourier basis and avoiding 
the variety of known weighting schemes, corrections, 
and assumptions -- and the differences in the 
data used.
This figure compares  
the average of our three x, y and z projected spectra 
in Figure \ref{fig_power_1}
with results from detailed analysis of very similar data by
\cite{tegmark2004b}
and of a much larger sample by \cite{PNE2007}.

Using a flux limited sample,
instead of our more easily interpreted
volume limited sample,
the first authors address the selection function,
redshift space distortions, bias effects
and other systematic errors,
using a Pseudo-Karhunen-Loeve expansion
\citep{tegmark2004b}. 
Figure \ref{fig_power_2} 
includes the data
from the first two columns of their Table 2 
in the form of open circles, 
without showing their rather large horizontal 
and vertical error bars.
They refer to this as the real-space 
galaxy-galaxy power spectrum $P_{gg}$ 
in units of $(h^{-1} Mpc)^3$,
and 
``recommend using column 2 for basic analysis."
Like ours this estimate treats the galaxies 
as equal points and accordingly is not corrected for bias,
justified because bias appears to be largely 
luminosity and scale independent (their Figures 28 and 29,
renormalizing to the linear $\Lambda$CDM model).
It is noteworthy that their power spectra
with and without correction for the 
Fingers of God (FOG)  -- their columns 2 and 3,
respectively --  would be indistinguishable 
had we plotted both.
Even though the effect of FOGs seems
insignificant here, 
redshift space distortions should be addressed
in any serious scientific applications.
Analysis of a much larger redshift survey,
extending to much larger redshifts than our study,
by \cite{alam2016}, 
includes significant redshift-space distortion corrections.
The results of a similarly detailed analysis by
\cite{PNE2007}
of a sample including both 
SDSS main galaxies
and  luminous red galaxies (LRGs) 
out to much larger redshifts
($z \sim 0.5$) than our sample, 
are plotted as plus-signs.

First compare the curves for the direct 
and binned transforms (solid and dot-dash
lines).  While the values at some
frequencies, especially the lower ones, differ by 
nearly an order of magnitude, the rough 
similarity of the slopes and values at higher
frequencies demonstrates
that these two
methods are crudely consistent with each other.
As well, the similarity of some of the finer detail 
in the two representations support the notion
that the effective spatial resolution is relatively good
(probably better than that corresponding to Tegmark
et al.'s horizontal error bars, not shown here).
The differences between our spectra and the others, 
especially in the form of a vertical offset above about 
$k \approx .05$, are not surprising in view of the
differences in the data and methods used.

Nominally the plotted points in our power spectra 
 are independent of each other.  
 Essentially no significant measurement errors propagate into this plot at any spatial frequency.
The only large discrepancy
in the plot is between  Percival et al.'s
and our powers at the longest scales,
understandable in terms of the difference
of the data samples and systematic 
effects at large scales. 
In the power spectra 
in \cite{tegmark2004b} and ours,
not surprisingly 
there is no evidence for 
baryon acoustic oscillation features.
These important features do 
begin to appear 
at around $k = 0.7 Mpc^{-1}$
with the larger  sample and the inclusion of the SDSS luminous red galaxies 
in \citep{PNE2007}.

A further avenue of investigation 
would invoke surrogate redshift surveys,
such as galaxy catalogs 
derived from $N$-body simulations.
The idea would be to compare 
simulated against actual distributions 
of variously grouped
power or phase spectrum quantities.
Using ensembles of synthetic catalogs 
to enable variance analysis 
is probably a fruitful 
path to reliable scientific conclusions.
While such a study is beyond the scope of this paper, we have carried out 
 a simple comparison 
against a single catalog, 
the Millennium Simulation (MS) of
\cite{springel2005}.
This $N$-body simulation 
(with $N = 2160^{3} = 1.0078 \times 10^{10}$)
contains data used 
to study multi-scale structure in Papers I and II.
The $xyz$ positions of simulated 
galaxies were treated in the same way as 
the SDSS data -- yielding a volume limited sample,
and without discarding galaxies lying 
close to the edges of the data space.
This MS analysis is described in Section \ref{error_analysis}.

\subsection{Gaussianity}
\label{non_gaussianity}

In modern cosmology it is often 
posited that the initial 
conditions of the Universe 
consisted of a random density distribution described as 
a \emph{Gaussian random field}.
It is not completely clear how 
the character of such an initial distribution 
may have evolved gravitationally,
or how matter-to-galaxy biasing,
integrated Sachs-Wolfe (ISW) effects,
and gravitational lensing 
may complicate conclusions based directly on
the galaxy distribution \citep{coles2000}.
Hence the interpretation of 
detected non-Gaussianity (NG) in the distribution 
of low redshift galaxies would not be straightforward.
We find no NG signatures here,
but if significant detections were to be made,
e.g. in future large redshift surveys, 
the resulting parameters would be useful
as additional constraints on precise 
cosmological evolution models.
Hence we now describe some aspects
of direct analysis of the Fourier phase spectrum.

Although Gaussian processes 
are well understood mathematically,
the elusive nature of 
\emph{non-Gaussian}  (NG) processes
has complicated 
and discouraged exploration of searches for their signatures.
The infinite number of ways 
a process can depart from Gaussianity
leads to a plethora of potential NG metrics,
only a handful of which have been pursued.
Here we describe a relatively straightforward
class of NG tests based on metrics of Gaussianity 
applied to the complex Fourier data cube.
The idea centers around 
metrics of how identically and independently
(IID) the Fourier phases 
at different spatial frequencies are distributed.

Much previous work centers on parametric tests,
valid only in the context of hypothetical
physical or mathematical models 
and thus far short of 
general characterization of NG.
Analyses using higher-order spectra and correlation
functions, or function bases such
as Karhunen-Loeve expansions \citep{vogeley_szalay_96,tegmark2004b}
or harmonic oscillator eigenfunction expansions 
\citep{rocha2001}
are closer to the spirit of non-parametric
analysis with its greater generality and flexibility.
On the other hand these methods are
simply ad hoc ways to project 
an infinite dimensional function space into
lower dimensions for modeling convenience.
By contrast the approaches of \cite{rocha2001}
and \cite{contaldi2000}
employ Bayesian frameworks
that alleviate some of this ad hoc character. 
But the conclusions are still dependent
on the correctness of a hypothetical model 
(e.g. the quantum mechanical
harmonic oscillator in \cite{rocha2001}).
More recently \cite{kovacs2014} define 
\emph{generalized phases} and apply this concept
to characterize the coherence between
WMAP and Planck CMB maps.

In the CMB context 
various authors have made suggestions
for the two aspects of this problem,
namely identification of:
(a) phase subsets 
that are computationally practical  but 
do not discard too much information,
and 
(b) non-Gaussianity metrics for these sets
\citep{CNVM2003,CNC2004,NCON2005,CN2007}. 
For one example \cite{CNC2004} discuss 
a number of general problems 
and propose an innovative procedure using return maps.
This can be thought of as a way to quantitatively characterize 
joint distributions \cite[cf.][e.g.]{scargle_iv}.
In another example \cite{CN2007} 
propose ring-like sets in spatial frequency
space.
And more recently several authors have
proposed phase analysis based on 3-point 
correlation functions of the Fourier
transform of a whitened version of the density field 
\citep{WBO2015,EBSN2015}.

\subsubsection{The Fourier Phase Spectrum}
\label{phase_spectrum}

We utilize the Fourier transform 
as a convenient setting for quantitative characterization of the Gaussianity of the spatial distribution of galaxies.
Keep in mind that the histograms constructed
in pursuing this goal are simply 
distributions from 
the one data sample on hand -- not probability distributions
with respect to some stochastic ensemble.
Several considerations motivate our focus on phases:

\begin{enumerate}

\item[(1)] The phase spectrum captures much 
of the information on non-Gaussianity present in the data.

\item[(2)]  Phases of Gaussian data 
are identically and independently distributed
\citep[IID; see e.g.][]{NCON2005}.
Measures of dependency in the phase distribution 
are consequently measures of non-Gaussianity.

\item[(3)]  The oft used
bi-spectrum and higher
order spectra and correlation functions have many problems.

\item[(4)]  In most practical, largely non-astronomical, situations
 the Fourier phase information
in 2D images is much more important than
the amplitude information 
\citep{oppenheim1981,coles2000,ChiangC2000}.
See also 
\citep{mannell1990} for a related discussion
of phase in speech intelligibility.

\end{enumerate}
\noindent
In addition see comments in Section \ref{epilog}
regarding related statistical
methods to be pursued in future work.
\clearpage

Point number 3 deserves more explanation.
Several authors 
\citep[e.g.][]{ferreira_magueijo1997,carron2011,carron_neyrinck,carron2015}
have disclosed a variety of 
fundamental problems with the multi-point 
correlation function hierarchy,
including 
large computational complexity that grows rapidly with
order,
information 
mixing among the orders,
and the fact that even including all orders 
only incomplete information is captured for
a log-normal density field
\citep{carron_neyrinck}
and only a tiny dramatically decaying fraction 
of the total information content of large fluctuations 
\citep{carron2011}.
A direct  Fourier transform avoids to a large degree 
all of these problems:
computational complexity
is relatively small
($N \times $ the number of spatial frequencies),
the amplitudes at different spatial frequencies 
are independent of each other,
convergence is well understood,
and the invertibility of the Fourier transform (Appendix A) 
assures that the combination of the
power and phase spectra capture
all of the information in the data.
Most important, 
simple data cubes  
cleanly display phase as 
an explicit 
function of $\bf{k}$. 
Furthermore there are no
special problems like those with generating 
representative sets of triangles and avoiding 
oddly shaped ones, as 
for 3-point estimators.

\cite{coles2000} gives a clear 
discussion of the 
background for 
these points, to which we add 
only a few remarks.
Of course the basic notion
is that the Fourier Transform appraises 
structure as a function of scale.
The discrete estimate is a finite sampling of 
a potentially infinite number of degrees of freedom.
But the Nyquist-Shannon sampling theorem
guarantees that it captures 
all the information contained in the data, 
limited only by the data resolution.
Since the inverse Fourier transform
exactly recovers the raw data,
it is clear that the (frequency dependent) 
amplitudes and phases contain 
complementary information, 
together yielding  a complete description of the data.
The Fourier \emph{power spectrum}
completely characterizes the 
Gaussian properties of the data;
while non-Gaussianity information 
can appear in both amplitude
and phase spectra, 
in many situations the latter dominates.

Driven by these comments our basic approach is to 
study phase distributions for non-uniformity.
Perhaps the simplest possible 
approach is to examine the overall distribution 
of phases without regard to spatial frequency.
Any structure in this 
distribution would suggest the presence of 
underlying non-Gaussianity.
\begin{figure}[htb]
\includegraphics[scale=1]{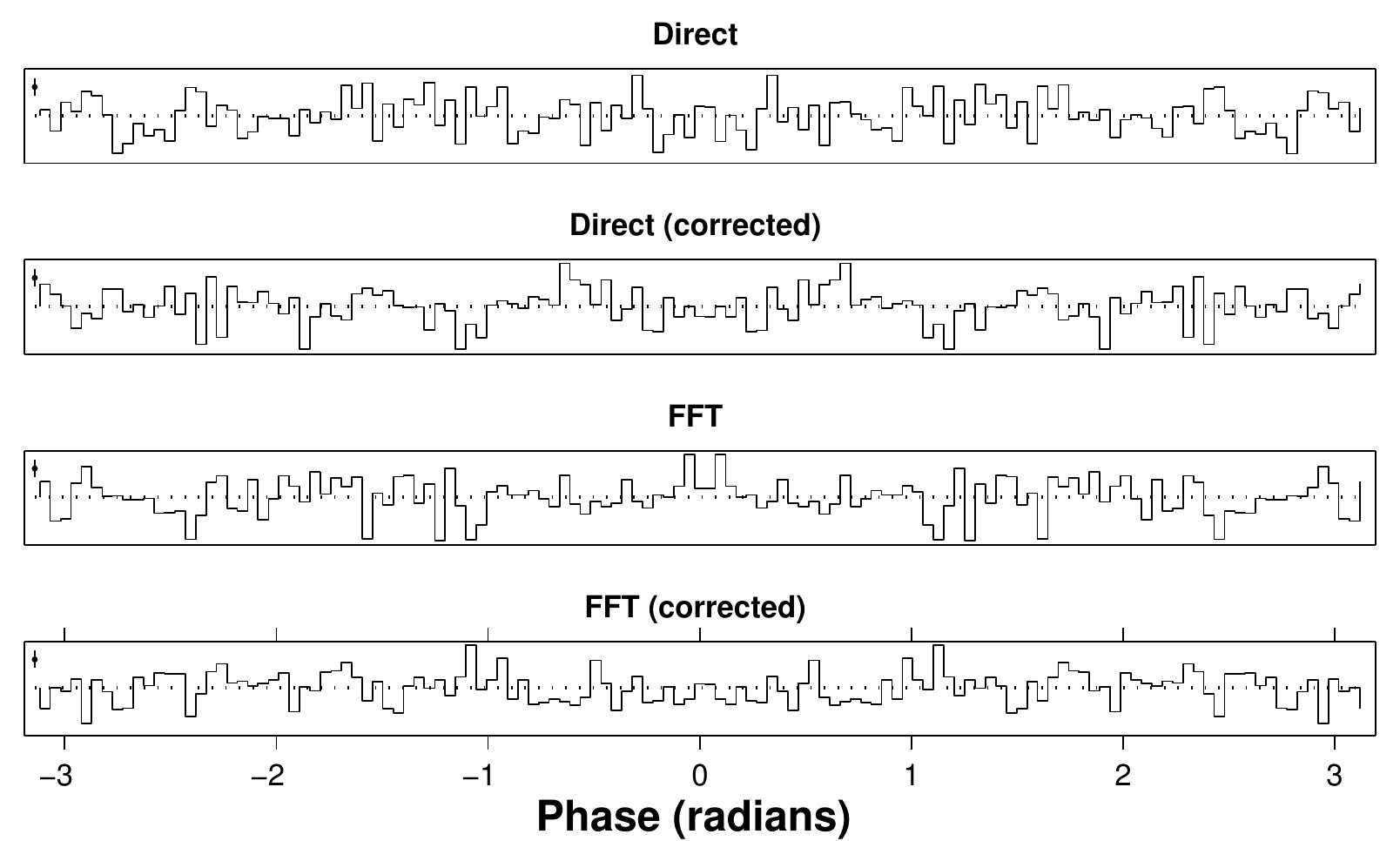}
\caption{Distributions of phases for the four cases 
(Direct as in eq. (\ref{ft_a}), and simple FFT of binned 
data as in Section \ref{3D_binned}, both with and without
correction for the window function, as labeled).
Horizontal axis is phase in radians.
The vertical axis is the 129-bin histogram population
of phases from the $128 \times 128 \times 128$ 3D phase cube, with a 
horizontal dotted line at the expected 
rate of $128 ^{3} / 2 \pi =  333,772.1$ counts per radian,
the ranges of these plots are $12,000$ is the same units. 
Poisson count error bars for a typical
bin are shown in the upper left corner of each plot.
}
\label{fig_phase_dist}
\end{figure}
Figure \ref{fig_phase_dist}
shows simple histograms of all 16 million-plus phases for the four cases, 
with 256+1 spatial frequencies in all 3 dimensions.
If these plots were scaled to include the zero of
the ordinate the fluctuations would be invisible.
There is here no evidence 
for any departure from uniformity,
but these overall distributions are almost certainly 
insensitive to NG because they do not take 
account of frequency relationships,
which are discussed in the next section.

It is perhaps notable 
that even the distributions for phases 
uncorrected for the data window
(the first and third panels) do not reveal 
perceptible nonuniformity.
This somewhat surprising result
probably reflects that the data window 
truncates the Fourier components
but does not change their phases.

\subsubsection{Distributions of Phases 
Grouped by Spatial Frequency}
\label{groups}

A more refined approach is to aggregate 
phases into two or more sets and test 
whether the distributions in them are identical,
as they should be in the Gaussian case.
The model-independent and non-parametric way 
the phase spectrum neatly lays out the
relevant information in a 3D data cube
facilitates such segmented analysis.
Accordingly we use the following
procedure to study NG:
\begin{description}\itemsep3pt
\item[(a)] Compute the complex 3D Fourier transform
$A(k) e^{ i \phi(k)}$.
\item[(b)] From (a) compute the 3D data cube 
$\phi(k_x, k_y, k_z)$.
\item[(c)] Specify a collection of subsets of (b) to be tested.
\item[(d)] Evaluate differences of 
nearest-mode phases within each member of the collection (c).
\item[(e)] Select an IID metric and compute it for each of the 
differenced arrays in (d).
\item[(f)] Assess the statistical significance
of the results of the collection of tests (e).
\end{description}
\noindent

The first two steps are straightforward
from the discussion in Section (\ref{3Dft}).
Step (c) addresses the need to 
identify sets of frequencies 
related to each other in some germane way.
For example 
phases at nearby frequencies would presumably 
show dependencies when those at well-separated 
frequencies might not.
From among the many possible ways to 
take advantage of the organized way 
frequencies are arranged in a  
3D phase-data cube, we adopt the following:
Let $N_{k}$ be the size
of the Fourier transform in each of its 3 dimensions.
For each of the $N_{k}^2$ pairs $(k_y, k_z)$,
for the array consisting of the 
corresponding $N_{k}$ phase 
values as a function of $k_x$ -- we call this array 
an \emph{$x$-beam} --
compute a metric or test statistic $T_{x}$,
and similarly for $y$-beams and $z$-beams.

Step (d) implements suggestions in
\citep{ChiangC2000,CC2000,CC2001,coles2000,wcm}
emphasizing the potential effectiveness of studying 
the differences between phases at adjacent spatial
frequency modes,
as opposed to the phases themselves.
Consider one dimensional first differences of the form
\begin{equation}
D_{k} = \phi_{k+1} - \phi_{k} \ 
\label{difference_1}
\end{equation}
\noindent
where $k$ is a spatial frequency index, 
here taken in one of the three 
cardinal directions -- x, y or z.
The following analysis was done with 
un-differenced, 
first-differenced, and higher-order-differenced
beams, the latter proving useful in unrelated 
sequential analysis problems (unpublished).
But un-differenced or higher-order differences
gave less clean results than 
first differences, 
so here we report only these results.

As noted by \cite{wcm} if the phases are IID
on the circle then so are their differences.
This requires that the phase differences 
that lie outside $(-\pi,\pi)$ first be adjusted 
for ``wrap around'' as follows:
\begin{eqnarray} 
D_{k}^{m} & \rightarrow  +2\pi - D_{k}^{m} \  \  \mbox{for} \ \ D_{k}^{m} > +\pi \\
     \  \ \          & \rightarrow -2\pi - D_{k}^{m} \ \  \mbox{for} \ \ D_{k}^{m} <  -\pi \nonumber 
\label{flatten}
\end{eqnarray}
\noindent
This procedure
is different from 
standard \emph{phase unwrapping} 
based on changing absolute jumps greater 
than $\pi$ to their $2\pi$ complement
(e.g. for the MatLab function \emph{unwrap}; 
\copyright \ the MathWorks Inc).

Step (e) 
amounts to the generation 
of a collection of estimates
of a NG metric of a beam in the form 
\begin{equation}
F_{x}( k_y, k_z ) = T_{x} \phi (k_x, k_y, k_z )
\ \
k_y = 1, 2, \dots , N_k; k_z = 1, 2, \dots N_k \ , \label{test_stat}
\end{equation}
\noindent
where the subscript 
just indicates which variable $T$ operates on.
At this point a huge range of 
possible definitions of $T$ opens up. 
There is no unique or universally best metric,
and undoubtedly different metrics are 
suitable for different forms of NG.
Item (2) in Section \ref{phase_spectrum}
motivates the use of a 
test statistic for the null hypothesis of IID phases.
The formal definition of independence
(joint distribution equals product of individual 
distributions)
is difficult 
to turn in to a practical IID test 
\citep[e.g.][and Section \ref{epilog} below]{scargle_i,coles2000,hyvarinen_book}.
Variance is a simple and easily 
interpretable statistic that might be sensitive
to some types of NG.
The Planck Collaboration 
studied both \emph{skewness} 
and \emph{kurtosis} \citep{ade2014}
-- measures of asymmetry of 
a distribution and of 
the relative importance of the 
center versus tails.
\cite{jin2005},
based on detailed study of various methods 
of CMB non-Gaussianity detection,
concluded that analysis of 
the kurtosis of wavelet coefficients is best.
Based on an idea in \cite{poly}, 
\cite{ChiangC2000} propose use of  \emph{phase entropy}
for NG studies.
\cite{hyvarinen_book} claim 
the optimality of entropy as an NG metric,
but in practice use kurtosis 
as an approximation 
because of pitfalls in entropy estimation.
Skewness did not seem to 
add any NG detective efficiency 
compared to kurtosis,
so here we report studies of variance, 
kurtosis, and entropy.
In every case these metrics 
were applied to first differences
between phases at adjacent  
frequencies.

\clearpage

\subsubsection{NG Metric Maps: Control Samples}

Consider now the analysis of synthetic
NG-free data for a sequence of three 
increasingly realistic sampling schemes,
followed by analysis of the actual data.
Maps of the 
beam metrics defined in Eq. (\ref{test_stat}) 
are presented as images with 
greyscales of the metric defined along one dimension, 
as a function of the two perpendicular
dimensions in the phase data cube.
Within each of the following figures 
the three panels present the analysis of the same data
for beams in the x, y and z directions.
When the data fall within the irregularly shaped volume 
we have used different spatial frequency arrays 
in the three directions.
That is, the frequencies are
integer multiples of a fundamental frequency
defined by
\begin{equation}
k_{0} (n) = { 2 \pi \over L_{(n)} } \ \ 
\end{equation}
\noindent
where $L_{(n)}$ is the range of the data in 
coordinate direction $n$.
This choice gives slightly 
better deconvolutions.
In contrast, for convenience 
the power spectra presented above in Section \ref{3Dft}
refer to the same frequency array in all directions,
namely corresponding to the largest of 
$L_{x}, L_{y},$ and $ L_{z}$.

\begin{figure}[htb]
\includegraphics[scale=1]{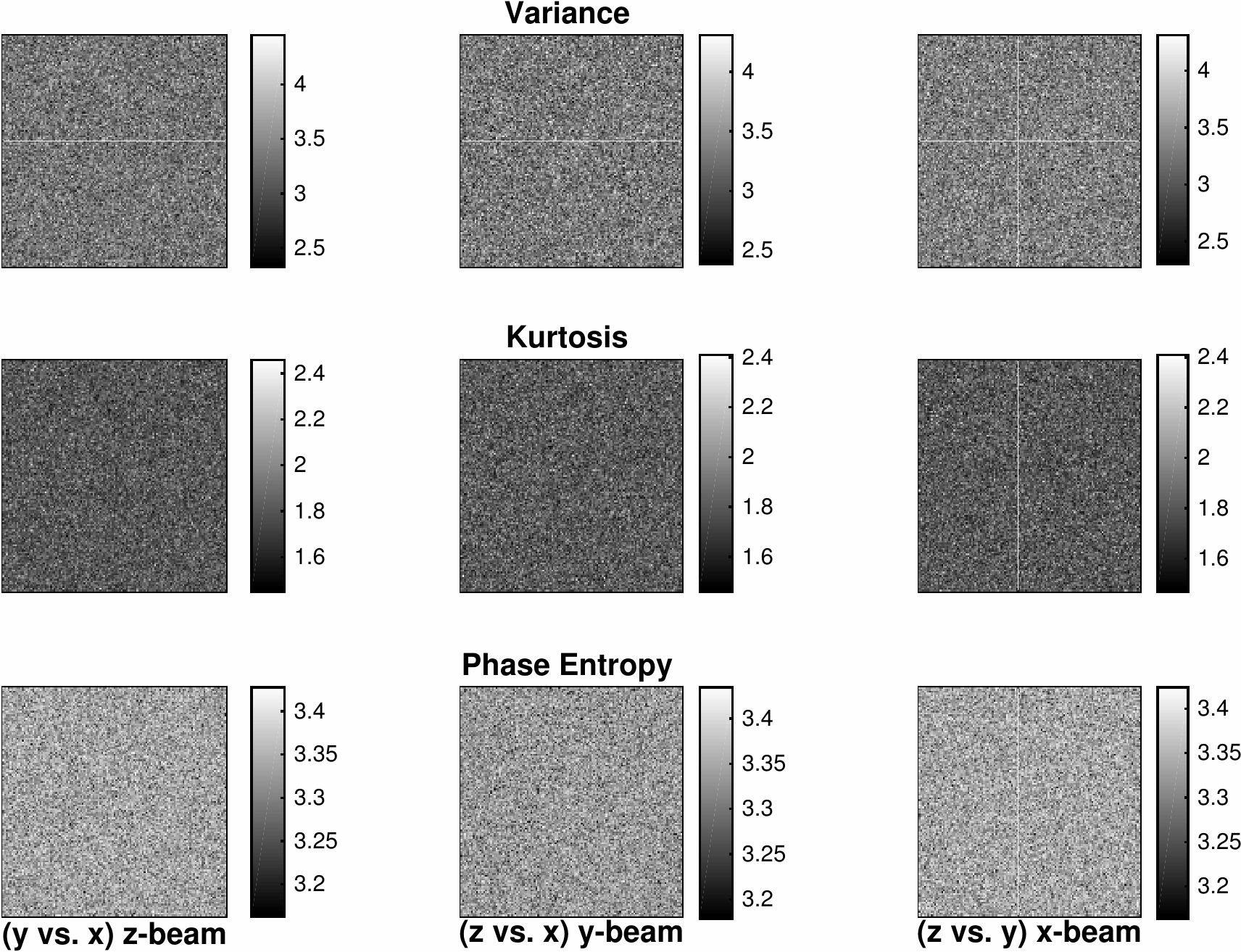}
\caption{Maps of NG metrics for random phases.
All images are from the same 3D 
$128 \times 128 \times128 $ cube of data 
consisting of IID random numbers 
uniformly distributed on $(0,2\pi)$.
Columns from left to right: beams in the x, y and z directions.
Rows (top to bottom):
variance, kurtosis and phase entropy.
Coordinates are indices in the 
synthetic random arrays,
not functions of spatial frequency as such, 
so axis labels are suppressed.
Here and in subsequent figures
the grayscale bars to the right of each panel 
depict the range of the metric.}
\label{phase_rand}
\end{figure}

\vskip 0.5in
Figure \ref{phase_rand} contains 
maps generated from a single 3D random phase cube.
Such images are 
used throughout this section to 
visually search for possible
non-random patterns in 
the behavior of NG metrics 
computed along beams parallel
to the coordinate axes.
The rows of panels are for the three metrics
-- variance, kurtosis and phase entropy  --
applied to the first differences of the phases 
adjusted as in Eq. (\ref{flatten}).
The columns are for 
the statistics computed along the x, y and z
directions.
These unsmoothed\footnote{Specifically we used the MatLab
$\mathtt{flat}$ mode for shading plots, not the 
interpolation mode
$\mathtt{interp}$.} 
plots retain the discrete 
nature of the data
to allow better appreciation of the randomness
of the distributions.
The data cube generating this figure 
consists of phases generated directly from 
an IID random number generator,
not from a Fourier transform, and thus
represents the simplest and 
most extreme form of the 
null hypothesis of IID phases.
As expected there is no apparent structure in any of
these panels.
\clearpage

\clearpage

\begin{figure}[htb]
\includegraphics[scale=1]{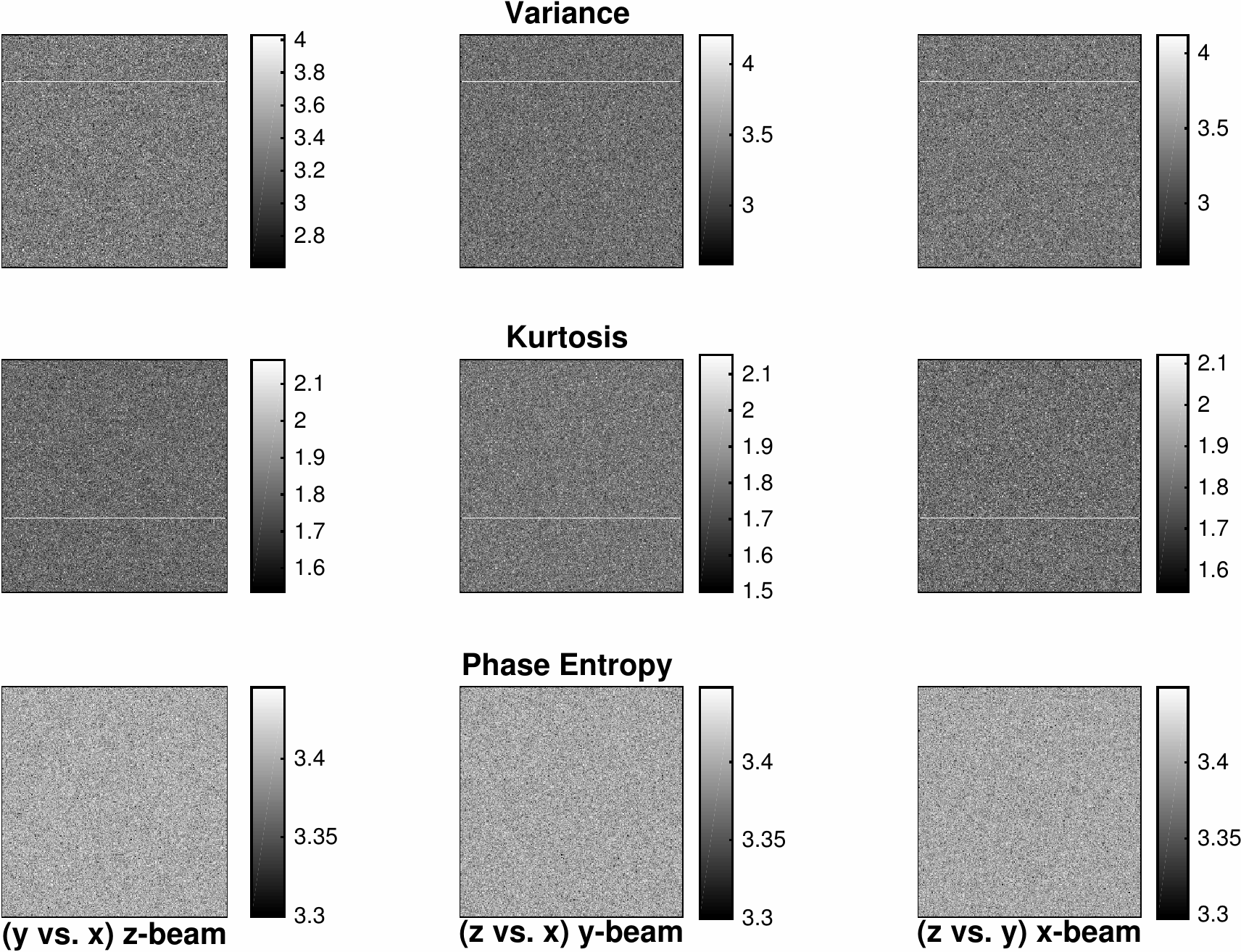}
\caption{NG statistics maps
for phases of the
direct Fourier transform of a set of
$100,000$ xyz points 
randomly and uniformly distributed 
within a cubic 3D volume,
using equation (\ref{ft_b}).
The identities of the panels are 
as in Figure \ref{phase_rand}.
While the coordinates are now spatial frequencies,
the units are fixed by the arbitrary size of the
cube, and therefore are also arbitrary.
The $128$ frequencies shown here 
cover the range $-f_{\mbox{0}} $ 
to
$f_{\mbox{0}} $,
where 
$f_{\mbox{0}} $ = $2 \pi / L$ 
is the \emph{fundamental frequency} 
and $L$ is the cube size;
zero frequency is the point at the
very center of the plot,
as in all the subsequent figures.}
\label{phase_rand_cube}
\end{figure}

Figure \ref{phase_rand_cube} 
shows the same type of map as
in Figure \ref{phase_rand}
but here the phases are derived
from the direct Fourier transform 
in Eq. (\ref{ft_b}) 
of random points distributed uniformly
within a $xyz$ cube.  
This configuration is chosen 
to diagnose possible modification
of the phase distribution inherent
in the transform procedure, 
but with a window that is benign 
due to the simplicity of its boundaries.
Deconvolution of the data window is 
not relevant,
due to the simplicity of the boundaries 
of the data space.
As expected there is no apparent structure in any of
metrics presented in these panels.

\clearpage

\begin{figure}[htb]
\includegraphics[scale=1]{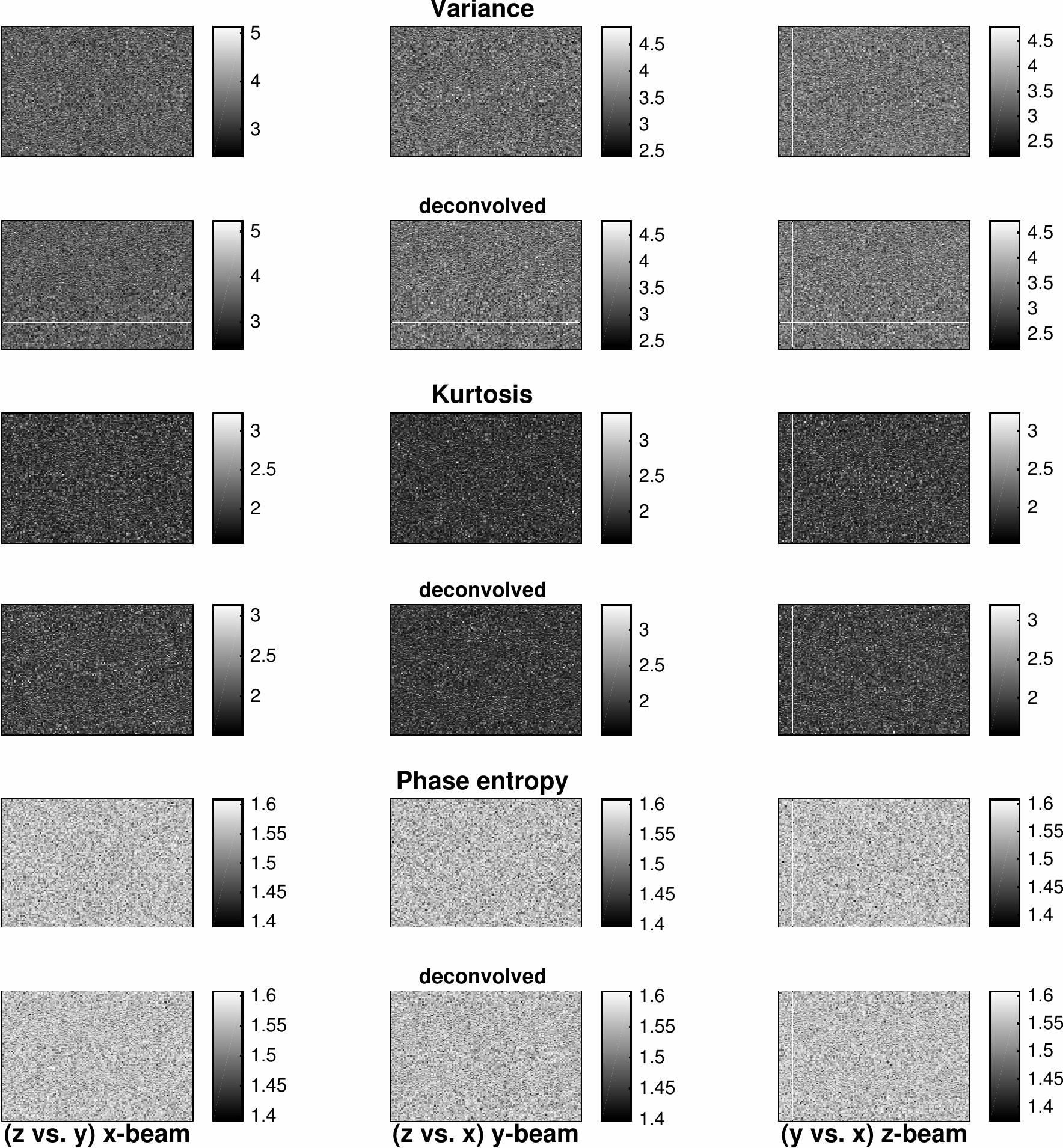}
\caption{Variance (first two rows), 
kurtosis (middle pair of rows) and phase entropy 
(last two rows) 
maps for phases 
from the Fourier transform of 
139,798 xyz points (the same
as the number of galaxies in our SDSS
data set) randomly
distributed within the convex hull of the
actual data.
The members of each of these pairs
are without and with data window 
deconvolution, respectively.
Columns are the 3 projections
as in previous figures.
}
\label{phase_rand_hull}
\end{figure}

Figure (\ref{phase_rand_hull})  
presents phase analysis 
for xyz-data that are still synthetic random 
points but now distributed uniformly
within the convex hull of the actual data.
The idea is to diagnose possible structure 
in these maps induced by the irregular 
boundaries of the data space.
The lack of structure here indicates 
that such distortion is minimal.

\clearpage

\subsubsection{NG Metric Maps: The Galaxy Data}

\begin{figure}[htb]
\includegraphics[scale=1]{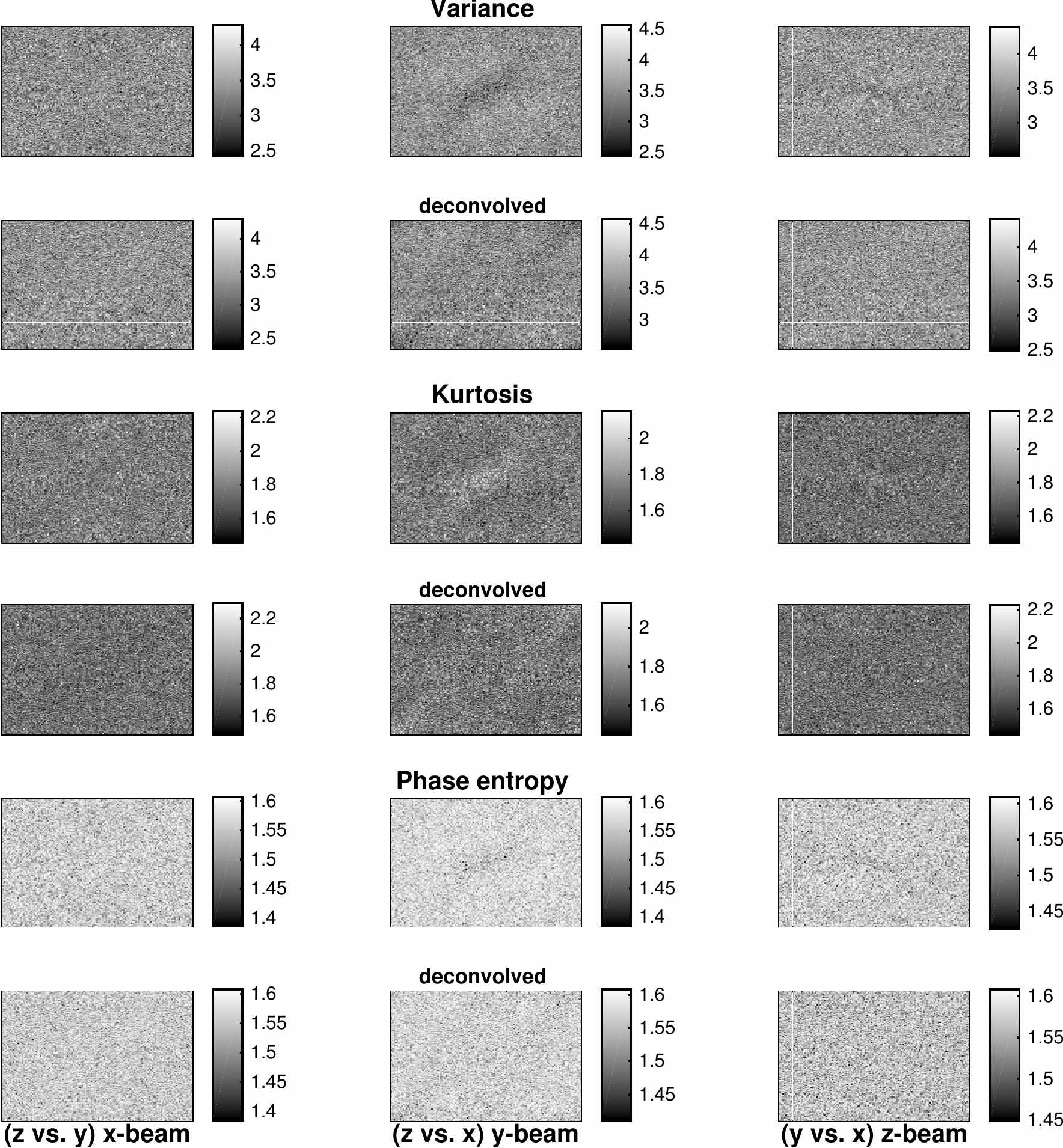}
\caption{NG maps
from the Fourier transform of 
the actual 139,798 galaxy positions,
displayed 
as in Fig. \ref{phase_rand_hull}.
Note that the kurtosis structure 
is lighter than average, as opposed
to the darker than average features 
in the other two cases.
The centers of the linear scales of 128 frequencies 
are 0; the adjacent points are $\pm 0.002 \mbox{Mpc}^{-1}$;
the ends of the scale are $\pm 0.128 \mbox{Mpc}^{-1}$.
}
\label{phase_xyz}
\end{figure}

Now turn to the actual galaxy data.
Figure \ref{phase_xyz}
presents the same analysis as carried out 
for the last of the 
control cases in the previous section.
The maps in the first, third and fifth rows,
depicting the statistics for the phase cube
not corrected for the data window, 
show clear evidence of structure -- most 
evident for the y-beam in the middle row.

\clearpage

The null results with the control samples 
in the previous section 
suggest that this  structure 
is not due to the irregularity
of the data window alone -- 
no non-random structure is evident 
in Figure \ref{phase_rand_hull} -- but
rather to the
combination of both the multi-scaled clustering in the 
point distribution and the irregular shape
of the data window.
In any case, 
the structure in all three beams 
(rows 2, 4, and 6) largely disappears 
when the data window has been deconvolved
in the spatial frequency domain.

The subtle residual structure is possibly real,
but more likely reflects 
imperfect deconvolution 
and is therefore not of astrophysical interest.
This conclusion is reinforced by the
similarity of the morphologies of the
uncorrected and residual structure.
Figure \ref{phase_xyz} is simply illustrative of an approach
to a difficult scientific problem -- perhaps useful 
in future studies with larger data sets -- 
and  is of course not a definitive comparison 
of the three statistics
nor meant to imply that kurtosis 
or variance are superior metrics.
Indeed \cite{hyvarinen_book}
present evidence that 
entropy may be an optimal Gaussianity detector
(cf. Section \ref{epilog}).



\subsubsection{Phase non-Gaussianity due to Density Perturbations}

Under gravitational evolution 
from even a perfectly Gaussian initial state
the low-redshift galaxy distribution 
is likely to have developed 
some degree of departure from Gaussianity.
Hence more realistic tests of 
NG detection methodologies 
would involve simulated density perturbations.
This section explores the connection between 
the distribution of phase differences and 
nonlinear clustering  \cite[cf.][]{wcm}.

\begin{figure}[htb]
\includegraphics[scale=.85]{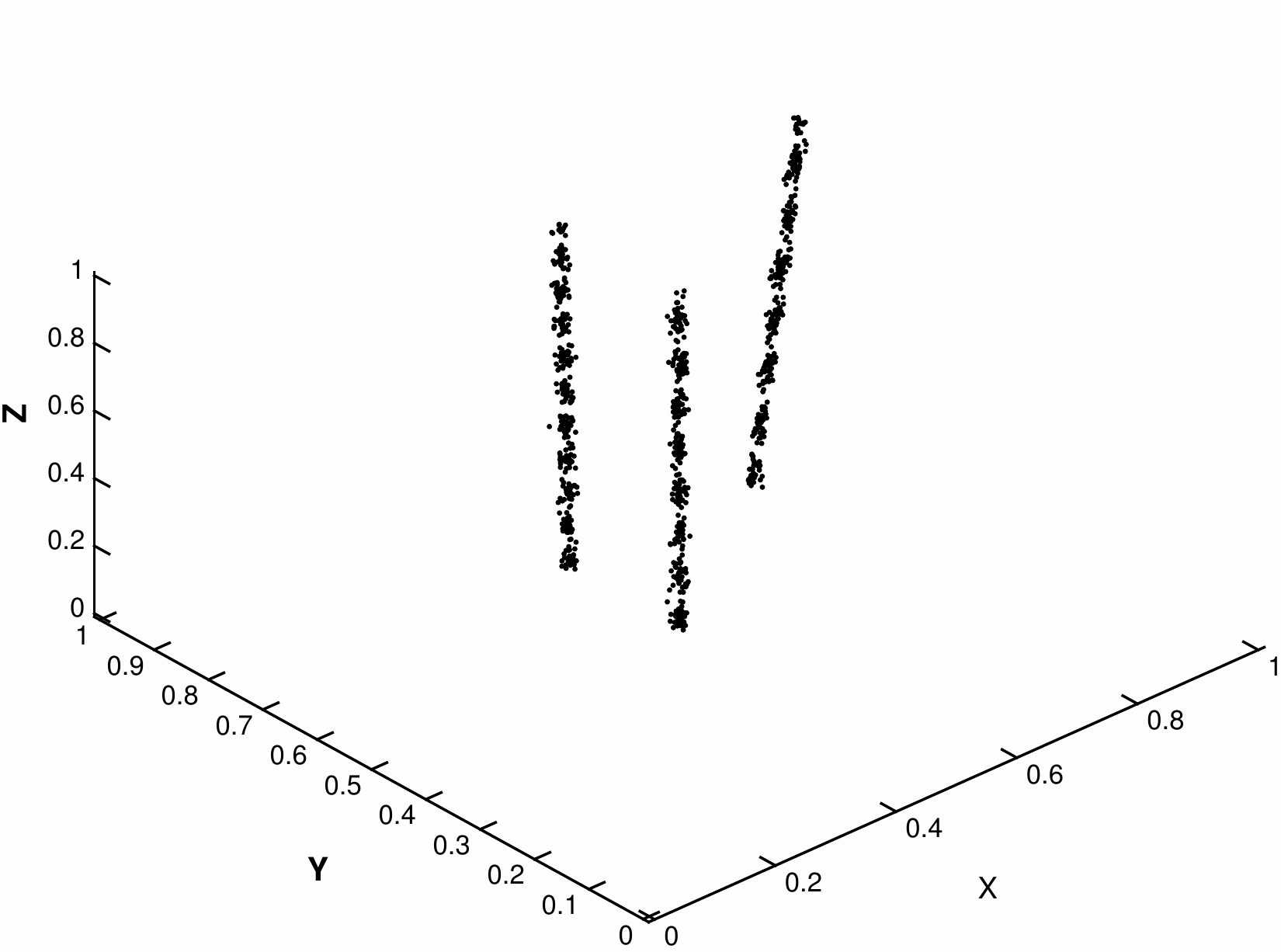}
\centering
\caption{Density perturbations
inserted into a unit 3D data cube.
Coordinates of the end points of the  three cylinders:
\#1: (0.5,0.5,0.0) - (0.5,0.50,1.0);
\#2: (0.5,0.7,0.0) - (0.5,0.72,1.0);
\#3: (0.8,0.7,0.0) - (0.9,0.72,1.0).
Transversely within each cylinder 
the points have a normal distribution of
standard deviation $0.005$.
The longitudinal density modulations 
correspond to sinusoids $1.1 + sin( kz )$ 
with $k = 50, 64$ and $45$
-- i.e. approximate periods of 0.12, 0.10 and 0.14 units. 
For visual clarity only $1,000$ points per beam 
are shown;
many more were used in the simulations as
indicated in the figure captions below.
}
\label{cylinders}
\end{figure}

We generated synthetic data 
consisting of points randomly distributed
in cylinders superimposed on a uniform background
within a unit cube.
These structures are not meant to be 
realistic models, for example of cosmic strings
or other topological defects;
they are constructed as pseudo-acoustic, 
transversely confined waves to introduce 
some degree of disturbance to the 
Fourier phases.
Each cylinder contains 
points drawn randomly from normal distributions 
with variance $0.005$ in
the transverse directions (x and y), and 
proportional to 
$1.1 + \mbox{sin}( kz )$
longitudinally.
Fig. \ref{cylinders}
depicts these cylinders: one is parallel to 
the z-axis; a second is slightly tipped ($\approx 1 ^{\circ}$)
and the third even more so ($\approx 6 ^{\circ}$)).
\clearpage

\begin{figure}[htb]
\includegraphics[scale=1]{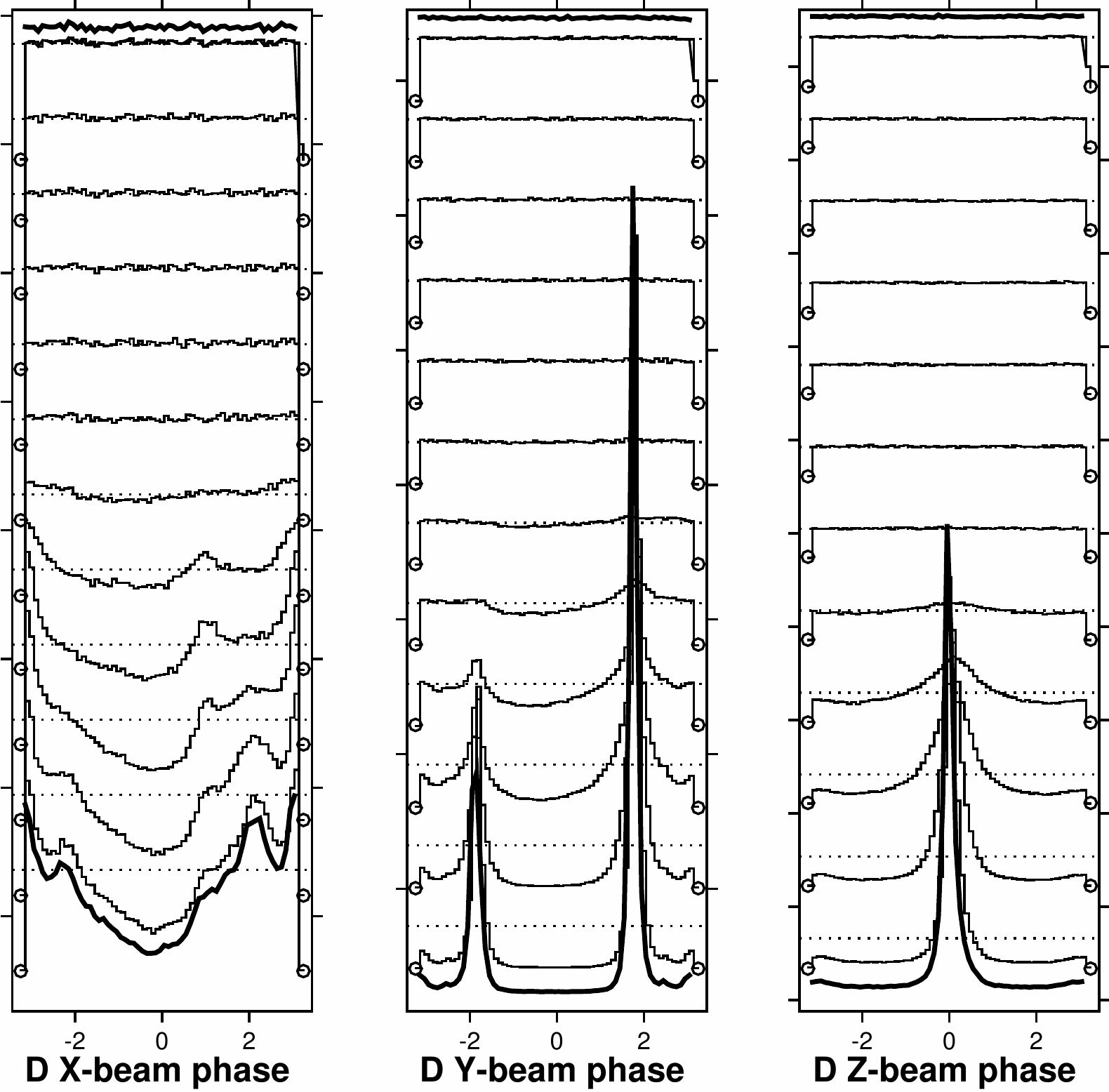}
\caption{Normalized 
distributions of 
nearest mode phase differences 
for random points in a 3D data cube
with various numbers of points 
drawn from the cylindrical configuration
of Figure \ref{cylinders}.
The twelve thin lines represent 
the distribution for the following numbers of
points in each cylinder:
$10, 32, 100, 317, 1000, 3163, 10000, 31623, 100000, 316228, 1000000, 3162278$,
against a uniform background of $10,000,000$ uniformly distributed points.
The thicker curves are for background only (top) 
and no background (bottom, $3162278$ points per cylinder).
The curves are shifted vertically for clarity; 
mean and zero levels are indicated by horizontal 
dotted lines
and circles at the curve endpoints, respectively.
Compare with 
Fig. 1 of \cite{wcm}.
}
\label{dt_1_figure}
\end{figure}

Realizations of this configuration
were superimposed on a dense random background of
uniformly distributed points, the former representing 
a perturbation of the latter.
The columns of Fig. (\ref{dt_1_figure})
display distributions of phase differences
\begin{eqnarray}
D( k_{x} | k_{y}, k_{z} ) = \phi_{k+1, k_{y}, k_{z} } - 
\phi_{k, k_{y}, k_{z} } \nonumber \\
D( k_{y} | k_{x}, k_{z} ) = \phi_{k_{x},k+1, k_{z} } - 
\phi_{k_{x}, k, k_{z} }  \\
D( k_{z} | k_{x}, k_{y} ) = \phi_{k_{x}, k_{y}, k+1 } - 
 \phi_{k_{x}, k_{y}, k }\nonumber
\end{eqnarray}
\noindent
\noindent
in the three indicated directions.
From top to bottom
the vertical sequences exhibit the evolution
of the 
distribution with
increasing NG signal strength.
As expected, for weak signals the distributions are
flat, but as as the NG signal strength grows
the distributions are more and more distorted.
At the seventh case (out of 12),
with $10,000$ points in each of 3 cylinders,
the distortion starts to become clearly significant.

\begin{figure}[htb]
\includegraphics[scale=1]{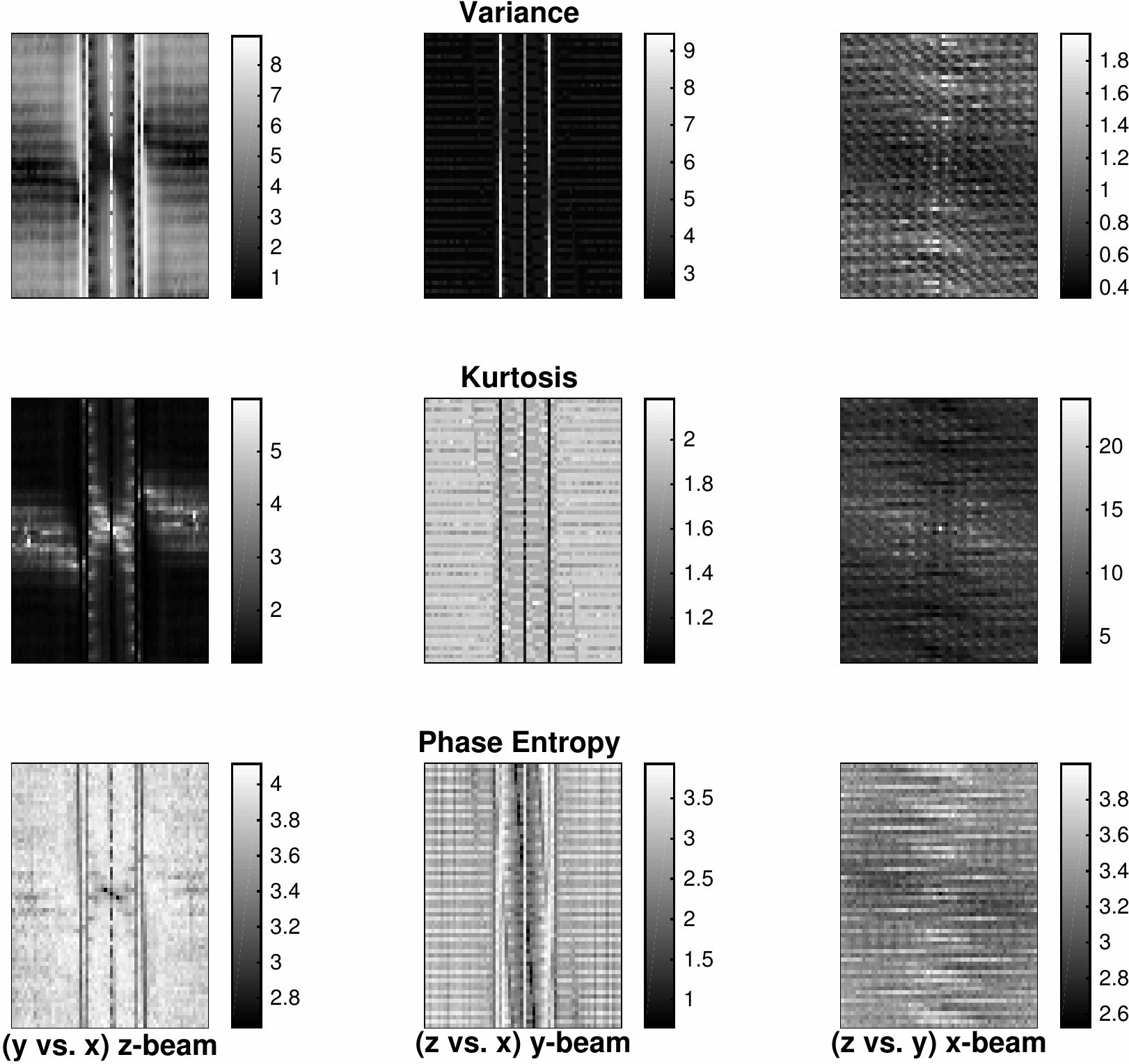}
\caption{Phase statistics maps for the toy 
three-cylinder density data described in Figure \ref{cylinders},
displayed
as in Figures 6-9.
The axis scales comprise 65 frequencies, with 
0 at the center;
the spatial periods corresponding to the
maximum $|k|$ are 0.0312 in units where
the cube edges are of length $1$.
}
\label{beam_14}
\end{figure}

Plotted in the same way as Figures 6-9, 
the maps in Figure (\ref{beam_14}) 
for the pure density signal 
(only points in the cylinders, no background), 
indicate that all three of these statistics 
reveal distinct spatial frequency  structure.
The maximum frequency in the Fourier
transform gives a minimum scale somewhat 
larger than the
approximate width of the cylinders;
therefore the
plots do not resolve these structures but reflect
the larger scales associated with distances 
between the cylinders, etc.

\clearpage

\begin{figure}[htb]
\includegraphics[scale=1]{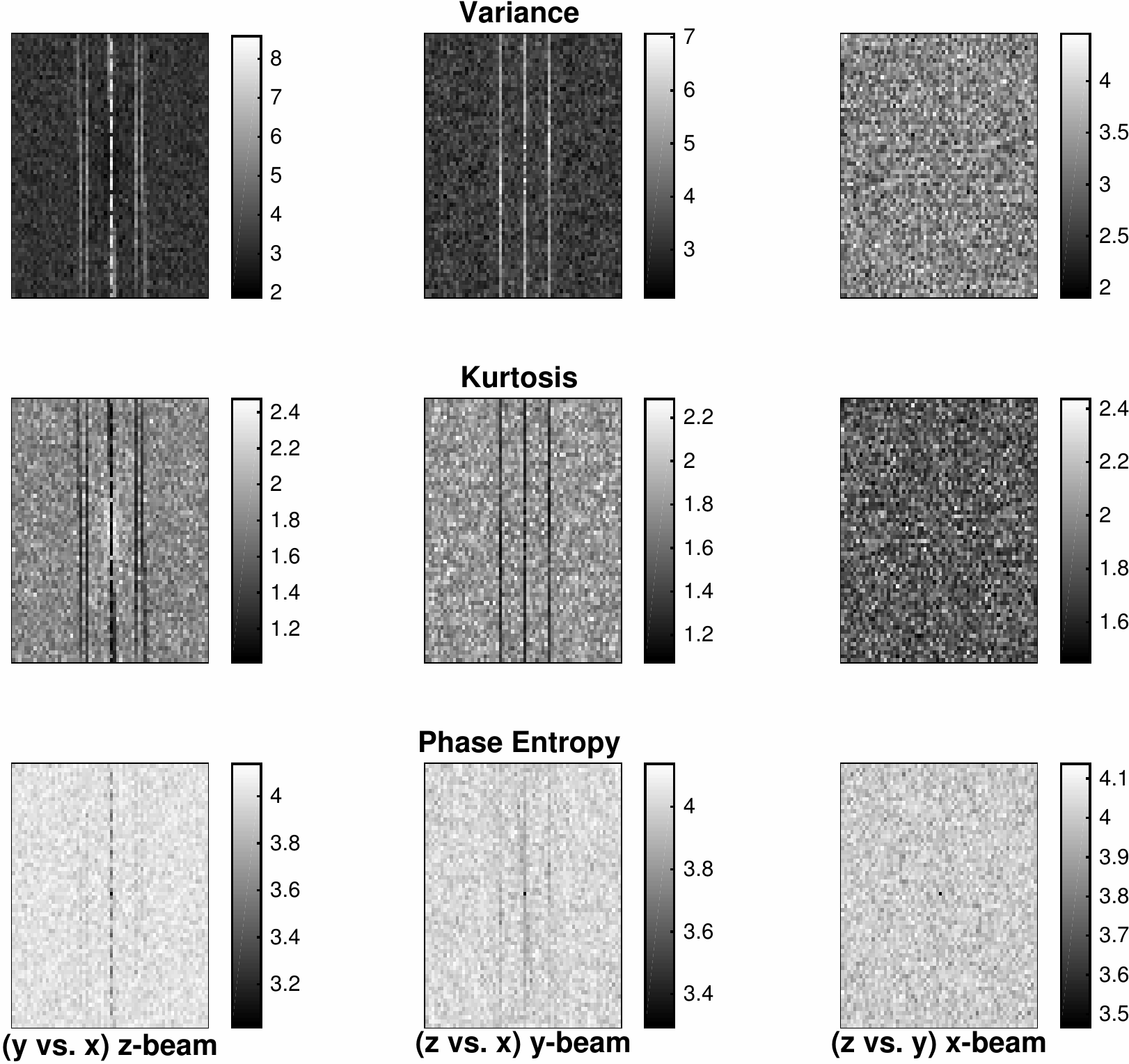}
\caption{Phase statistics maps for the toy 
three-cylinder density data described in Figure \ref{cylinders},
displayed as in Figures \ref{beam_14}.
This figure illustrates what a barely detectable
NG signature of the toy signal 
in Fig. \ref{cylinders} might look like, but of course is not a guide to realistic expectations.}
\label{beam_7}
\end{figure}

Figure (\ref{beam_7}) is for the case 
from  the sequence in 
Fig. (\ref{dt_1_figure}) where the 
weakest NG signal is just barely detectable
in the difference distribution:
$10,000$ points in each of the three cylinders,
with a background of $10,000,000$ uniformly
distributed points.
The three cylinders together thus contain the fraction
$0.003$ of the background.
The volumes of the cylinders are approximately
$\pi (2 \times .005 )^{2}$ --  smaller than the 
cube's volume
by a factor of $3 \times 10^{3}$ -- 
so the point density in the cylinders 
is about $3$ times 
the mean density of the background.

This section developed a visual approach
to assessing distributions of 
statistical parameters in a 3D data cube,
and applied it to try to detect 
departures from the hypothesis of IID Fourier phases.
In such displays 
the eye is famously good at perceiving patterns,
but also easily fooled by noise fluctuations.
Given  the display issues of 
pixelization, contrast, range, color, non-linear scaling, etc.,
and the difficulty of rigorous 
analysis of statistical significance
of perceived patterns in this kind of image,
a more objective
approach is called for, 
as addressed in the Epilog, Section \ref{epilog}.


\clearpage

\section{Uncertainty}
\label{error_analysis}

An estimate of the uncertainty 
of a scientific result 
is an important part of its value.
At issue is how widely the 
result might vary 
on account of the inevitable accidental aspects 
of the measurement process.
This can be addressed by appraising 
data values that could have been obtained 
but, by happenstance, were not.
Cosmology often sidesteps 
its one-universe handicap
by measuring uncertainty as the variance 
over a postulated distribution function 
of such hypothetical data.
In this section we discuss uncertainty
in our results using several 
ideas of what constitutes such 
``other data,''  in turn considering 
observational,
internal and external errors.

\subsection{Observational Errors}
\label{observational_errors}

One often has relatively good information 
about observational errors.
E.g. normal distributions 
with well-determined parameters
can often be theoretically justified
and empirically tested and calibrated.
Assessment of the corresponding
uncertainty is then relatively straightforward.
The only  sources of observational error 
relevant to our analysis 
are fiber collision effects
and random measurement errors in coordinates
and redshifts.
Paper II discussed our procedure for
mitigating the former, 
and we now demonstrate that the latter
are negligible.

\begin{figure}[htb]
\includegraphics[scale=1]{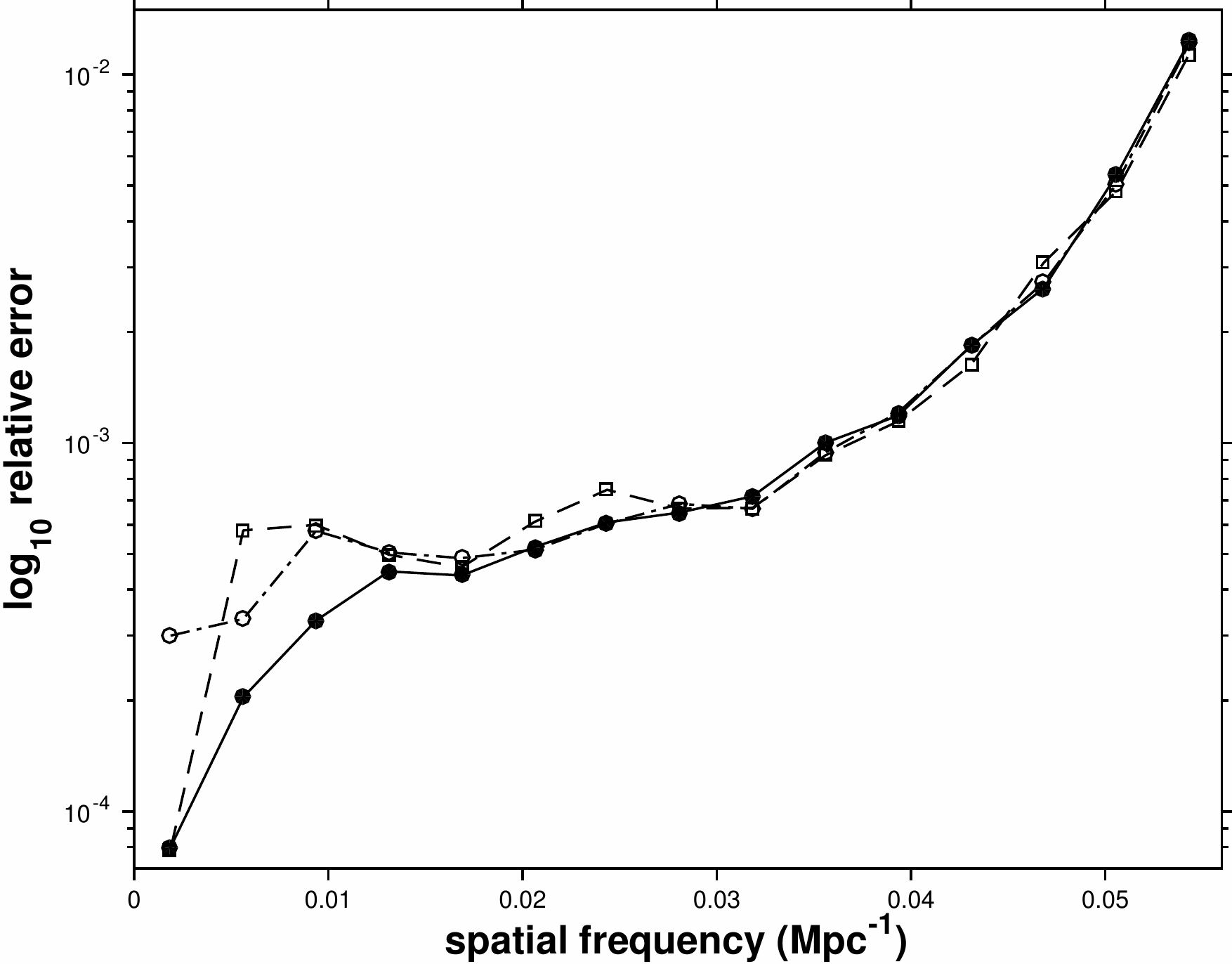}
\caption{Relative uncertainty from propagation
of the observational coordinate errors. 
The ratio of the standard error 
to the mean of the
power spectrum is plotted against 
spatial frequency. 
Solid line with dots, 
dashed line with squares,
and dot-dashed line with circles: power in the x, y, and z directions,
respectively.}
\label{fig_error}
\end{figure}

We simulated 100 realizations 
of normally distributed 
heteroscedastic errors
(zero mean and standard deviation as
given for each galaxy in the data catalog)
added to the actual 
right ascension, declination and redshift values.
Power spectra for these data sets were 
carried out exactly as for the actual data.
The relative errors were computed
as the standard deviations of the resulting
powers divided by the corresponding means.
Figure \ref{fig_error} plots these
results as functions of spatial frequency.
These relative errors are maximum
at the highest spatial frequencies,
reaching at most $1\%$.
Overall the effect of these errors is
at least several orders of magnitude
too small to have any relevance.
At the same time this analysis has 
ignored systematic errors and 
the likely possibility of correlated errors 
induced by systematics.
In principle a similar display of phase uncertainty 
is possible, but difficult to display and relatively 
uninformative, so we do not present it.

\clearpage

\subsection{Internal Variance}
\label{internal_variance}

An additional element of uncertainty 
arises because, not only could the measured galaxy coordinates
be different (as discussed in the previous 
subsection) 
but the sample could have actually contained
different galaxies.
The view is that the galaxy samples are randomly 
drawn from a hypothetical spatial distribution.
The relevant uncertainty 
is termed \emph{internal variance} -- that is, internal to the data space at hand.
One can think of this process 
as 3D spatial \emph{shot noise}.
This term typically refers to random fluctuations 
in a measured light curve of 
a varying astronomical source,
but 
here  the discreteness refers to galaxies
instead of photons.

One could compute variances in 
ensembles of random draws from 
a model of this distribution.
Shortcomings of any such model-based
approach include
loss of information 
by imperfect representation of the data,
imposition of incorrect information 
(e.g. by effective smoothing), 
and dependence on the correctness
of the model form and its parameter values.
Furthermore this procedure provides no evidence on 
what the distribution actually is.
Hence we do not choose to follow this approach.

Happily, random resampling techniques 
such as bootstrap and jackknife methods \citep{efron_tibshirani} enable straightforward 
model-independent estimation 
of internal variance.\footnote{
See also \citep{norberg2009}
for a related discussion of internal
vs. external errors
and comparison of various randomization methods
in the context of clustering statistics.}
Like most purportedly powerful and easily implemented
algorithms, these methods 
are sometimes misunderstood and used carelessly.
Two common reactions are that they are 
``useless; they seem to get something for nothing''
or ``great; they capture all relevant statistical information
from all kinds of data.''
The truth is in between but 
closer to the latter.
The following discussion 
shows what resampling can do here and what it can't.

The referenced resampling 
methods use the  empirical distribution function (EDF) 
to approximate the true distribution described above.
This function, derived directly from the data, 
captures all information contained therein about the true distribution.
Galaxy-by-galaxy resampling with replacement  or leave-one-out
cleanly implement the bootstrap or jackknife principle, respectively.
Resampling has the advantages 
that it relies on only the data measured,
needs no additional data,
and makes no assumptions other than that the 
empirical distribution is a good  
approximation of the actual one.

The \emph{jackknife} method uses 
a set of samples, each consisting of the 
full data set with 
with one point at random removed.
The \emph{bootstrap} method 
seeks the approximation mentioned above with
random draws from the EDF, 
defined to be the set of 3D coordinates 
$\{ \mbox{\bf x}_1, \mbox{\bf x}_2, \dots , \mbox{\bf x}_N \}$
each assigned the probability ${1 \over N}$, 
much as in Equation \ref{unit_sum}.
The result is simply a 
sample, typically $N$ in size,
randomly drawn  \emph{with replacement}
from the original data points.
That is, the randomly
drawn galaxies are not discarded
and may occur two or more times in the 
\emph{bootstrap sample}.
In both cases one simply analyzes 
many realizations of 
these surrogate data samples 
in the same way as the actual data.
The correctness of the results 
relies on the single assumption that the empirical distribution 
function fairly represents the underlying 
physical process.  
The \emph{bootstrap bias} or  \emph{jackknife bias} 
are estimates of any bias 
inherent to the  algorithm.
They compare the resampled mean against the original 
mean.
However they can say nothing about possible
bias in the original data themselves.

For the current power spectrum analysis 
the leave-one-out procedure of the
jackknife is almost trivial to implement; the $m$-th
jackknife sample, i.e. with the $m$-th point
left out, is from Eq. (\ref{ft_a}):
\begin{equation}
F_{\mbox{jackknife}}( \bf{k}, m) = 
  \sum_{n \ne m}^{N} \ 
    e^{ -i \bf{k} \cdot \bf{x_n} }  =
 F( \bf{k} ) -  e^{ -i \bf{k} \cdot \bf{x_m} } 
\label{jackknife}
\end{equation}
\noindent
Thus Fourier transforms of the jackknife 
samples can be computed without 
the need to evaluate the full $n$-sum 
each time.
Bootstrap samples are only slightly
more complicated.
These computational efficiencies 
allow the luxury of using $N$ resamples -- 
the maximum possible 
for jackknife and certainly overkill for bootstrap.\footnote{
Section 6.4 of \citep{efron_tibshirani} addresses the question of 
how many resamples are 
needed to assure good convergence.}

One more computational detail deserves mention:
The replacement aspect of bootstrap resampling 
yields the potential of 
algorithm problems with exact data point duplicates.
E.g. the local event rate measure
  $1 / (t_{n+1} - t_{n})$
in time series applications \cite[cf.][]{scargle_vi}
is infinite for duplicate event times.
Here there are no such singularities,
as the corresponding terms in the Fourier
sum simply add without difficulty.
Any concern is further alleviated 
since our bootstrap results 
are essentially 
identical to 
those using the jackknife, which 
does not generate duplicates.

\begin{figure}[htb]
\includegraphics[scale=1]{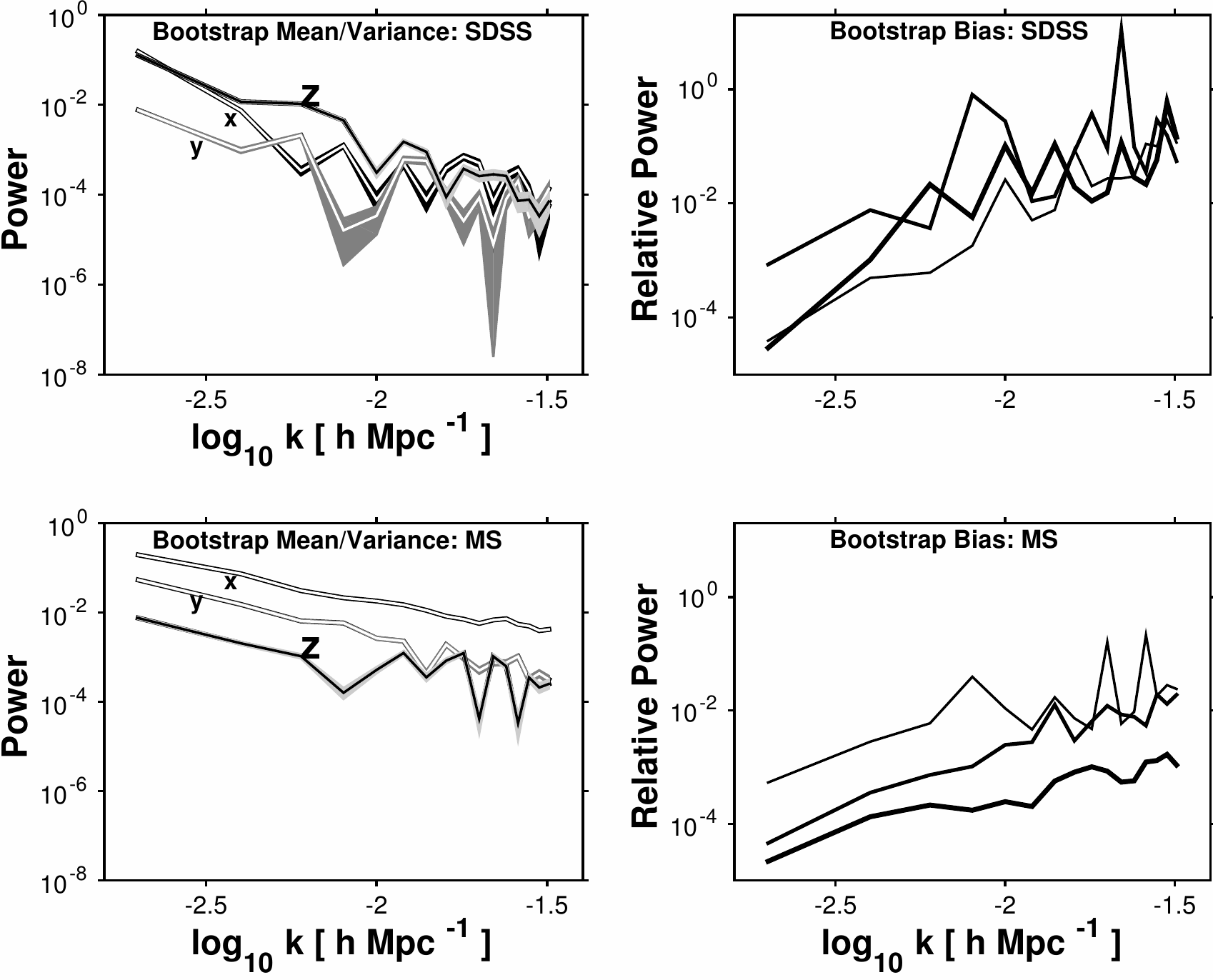}
\caption{Bootstrap mean, variance and bias of power spectra.
Left: x, y and z projections
of bootstrap mean power 
(in order of decreasing darkness and as labeled)
are plotted as narrow lines embedded in 
greyscale bands depicting the
$\pm 1\sigma$ bootstrap standard deviation.
Right: fractional bootstrap bias. 
Top panels: 139,798 bootstrap samples of the 
galaxy data.
Bottom panels: similarly 
for the Millennium Simulation data.
Jackknife results are indistinguishable from these.}
\label{boot_error}
\end{figure}

Reporting variance for a 3D spatial distribution 
is notoriously difficult; instead 
we choose to report it for power spectra. 
Figure (\ref{boot_error}) 
displays bootstrap means, variances and
biases for our standard SDSS galaxy sample.
It is clear that
the bootstrap components of internal variance 
and bias 
are quite small, especially at low spatial frequencies.
The bottom panels are similar plots for 
bootstrap analysis of the comparable
Millennium Simulation sample detailed 
in Papers I and II.  
Crudely speaking the
results are similar, although the
synthetic data yield somewhat less 
noisy power -- with smaller 
variance and bias.

\subsection{External Scatter: Cosmic Variance}
\label{external_variance}

Survey data from different regions of the Universe
give different parameter estimates.
Such 
\emph{cosmic variance}\footnote{The 
term is sometimes used in
other ways, but here is restricted to 
variance of parameter estimates 
over an ensemble of sub-volumes.} 
refers to errors in cosmological 
parameter estimates for scales
larger than covered by 
a given survey.
This uncertainty is in addition to 
those discussed in the above sub-sections.
To estimate the effect of cosmic variance 
on our results we Fourier analyzed 
ensembles of sub-sets of two data sets.

The first, 
drawn from the SDSS 
DR13\footnote{http://www.sdss.org/dr13} \citep{DR13},
is significantly larger than the original DR7 selection of Paper I.
It was obtained with a very similar query 
from the SDSS
skyserver casjobs web interface\footnote{https://skyserver.sdss.org/CasJobs},  
except that 
the absolute magnitudes were 
obtained directly in the casjobs query, whereas in Paper I we 
had to obtain this information via a cross match to the SDSS NYU VAGC
catalog\footnote{http://sdss.physics.nyu.edu/vagc} \citep{Blanton05}.
To further increase the size of the sample, while 
mostly remaining within the precepts of
the data selection of Paper I, we adopted
a somewhat larger redshift range ($0.005 \le z \le 0.15$),
and defined the volume limited sample
with a slightly fainter cut in absolute R magnitudes, at
 -19.8495 instead of -20.1 as in Paper I.  
 In addition we discarded 
 galaxies with anomalously large R magnitude errors,
 adopting a threshold of 0.2786.
 As in Paper I we selected only the contiguous north galactic cap 
region and applied the same procedure 
to address the fiber collision bias.
The resulting data set consists of $N = 370,847$
 galaxies in a (convex hull) volume of 
 $100.7 \times 10^{7} \mbox{mpc}^{3}$.

 The second data set is the same one 
 used in the current paper and defined in Paper I, 
 roughly $2.6$ 
 times fewer galaxies in a volume $10$ times smaller:
 $N = 139,798$ and 
 (convex hull) volume of  $9.4676 \times 10^{7} \mbox{mpc}^{3}$.
 In both cases these main data sets were 
 subdivided into 8 independent subsets:
 octants, with divisions 
 at the median values of the xyz coordinates.
 Summary statistics for these subsets appear 
 in Table \ref{cosmic_variance_table}.

\begin{figure}[htb]
\includegraphics[scale=1]{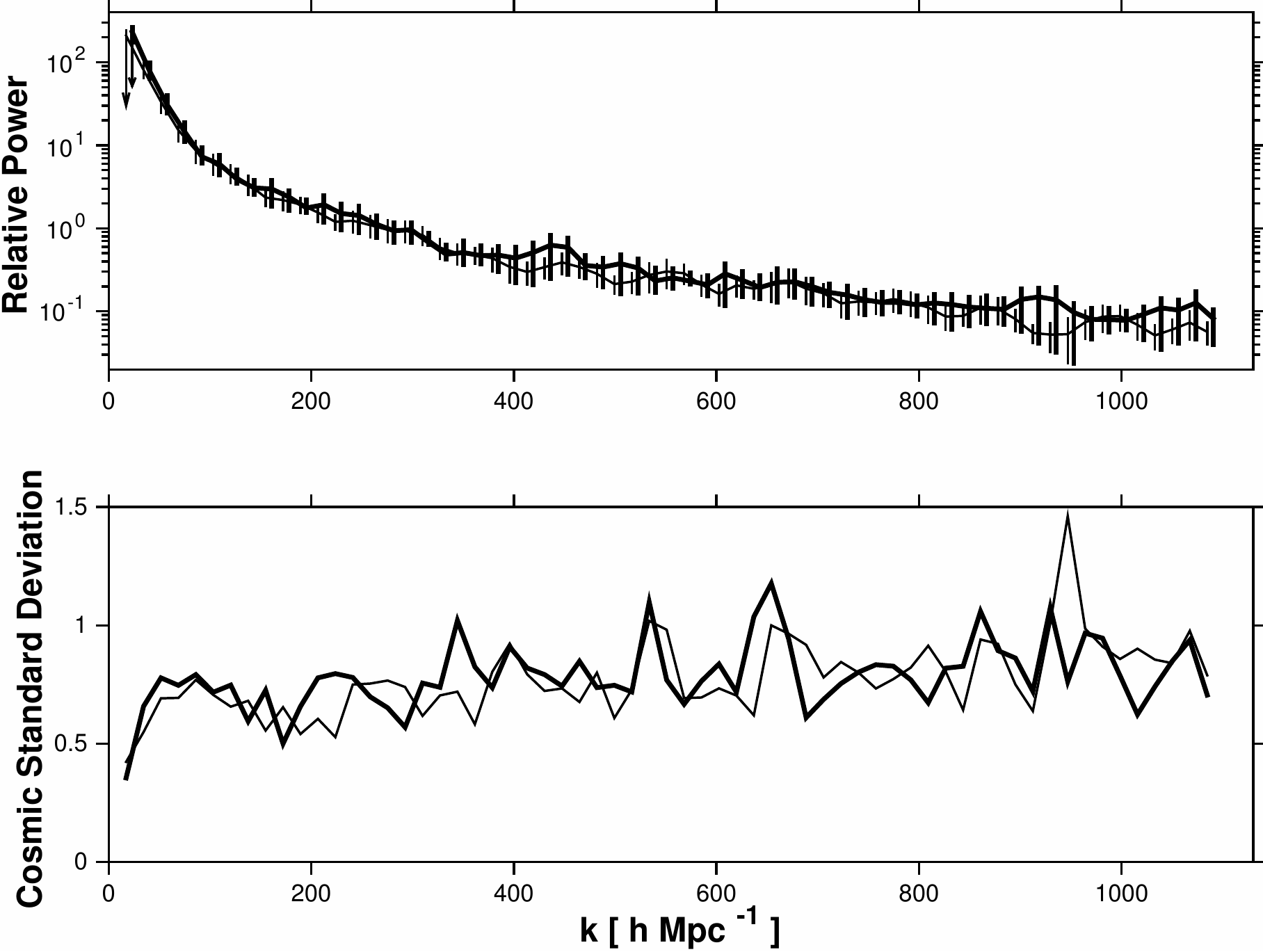}
\caption{Cosmic Variance. Top: Linear plots of mean
power spectra (averaged over 24 values:
8 octants $\times$  projections in the three
coordinate directions) and corresponding
standard deviations.
Bottom: the above standard deviations divided by the 
means, as a function of spatial frequency.
In both panels thin and thick lines are for
DR13 and DR7, respectively.
}
\label{cosmic_variance}
\end{figure}

The top panel of Figure \ref{cosmic_variance} 
shows linear plots of the power spectra,
averaged over the 8 octants (as well as the 
three coordinate projections).  The error
bars are the standard deviations,
which serve as estimates of cosmic
variance in the spatial power spectra.
The DR13 sample has somewhat
smaller scatter, as expected on account of
its larger size.
The relative size of these uncertainties
(standard deviation divided by mean)
is plotted as a function of spatial frequency
in the bottom panel.
The cosmic variance of the power values 
at a given frequency are rather large,
but one expects the more cosmologically
relevant 
\emph{normalizations} and \emph{logarithmic slopes }
of the spectra to be less uncertain, 
because they essentially average 
over this scatter as a function of frequency.

\clearpage

That this is the case can be seen in the information 
on these parameters in Table \ref{cv_table}.
The 8 numbered columns refer to the octants,
i.e. the independent sub-samples of the
survey data described above.
For the two data samples three quantities
are tabulated for each octant: 
the normalization\footnote{As noted in Section \ref{3D_direct} 
the value of the spatial power
at zero frequency simply reflects
the size of the sample.
Hence we chose to tabulate here 
values derived from the
intercept of the linear fits,
scaled to their mean.
These values are therefore not independent
of the other two tabulated data.}
and logarithmic slopes
from least-squares fits (linear in log-log space) 
to the spatial power spectra, and the number of
galaxies in the octant.
The next two columns 
give the corresponding means
and standard deviations -- first 
averaging the $x-$, $y-$ and $z-$ 
projections and then over the 8 octants.

We are interested in the uncertainties 
for the results derived in this paper for the 
full DR7 sample.  Accordingly,
the standard deviation values for the octants
in the penultimate column 
are adjusted
downward by the factor $\sqrt{8-1}$ 
to account for the relative sizes of the
full and sub- samples,\footnote{A straightforward 
Monte Carlo study validated this procedure, 
in spite of the fact that both the power spectra 
and their derived parameters are non-linear 
functionals of the data.}
and reported
in the last column (under the heading ``This Paper'').
The difference between the values for DR13 and DR7
data indicates the approximate
uncertainty of these determinations and
their extrapolation.
The fact that the percentage variance in normalization 
is smaller than in slope may be related to the 
comment in footnote 9.

It is useful to compare these results 
with the quantification of cosmic variance 
by \cite{driver}.
These authors studied the variance of galaxy density 
across independent subsets of 
much the same SDSS as used here. 
They derived approximate formulas 
for the corresponding standard deviation 
as a function 
the volume and aspect ratio of the 3D survey region,
and the number of independent sight lines.
Their Equation (1) 
gives the values reported 
in the second part of the last column,
for density cosmic variance for our full DR7 
sample.
The close agreement between
our $6.5\%$ (power spectrum slope) 
and their 
$7.0\%$ (galaxy density) 
for the average of the DR13 and DR7
extrapolations  is probably partly due
to the similarity of the data used in the
two works and partly fortuitous.


\begin{table}[htp]
\caption{Cosmic Variance: Power Spectra of Octants }
\label{cosmic_variance_table}
\begin{center}
\begin{tabular}{ | l  |  r |  r | r |  r |  r |  r  |  r | r || r |  r || l |  }
\hline 
Octant  & 1 & 2 & 3 & 4  &5  &6 & 7 & 8  & Mean & $\sigma$ 
& Full DR7 Estimate  \\
\hline\hline
SDSS DR13  & & & & & & & & & &  &This Paper \hskip 0.1in  Driver \\
\hline
Normalization &   1.170  &   0.912  &  1.071  &  0.988  &  1.012   & 1.020  &  0.911  &  0.988 &  1.009 & 0.084 & 3.0\%  \\
Slope &  -2.282  & -1.817  & -2.126  & -1.910  & -2.030 &  -2.022   & -1.823  & -1.953 & -1.995 & 0.157 &  5.6\% \hskip 0.6in  8.1\% \\
N (370,847) & 47181  & 40143   & 50153  &  47946  &   49092  &   49005  &     38995  &   48329 & 46356 & 4290 & \\
\hline\hline
SDSS DR7 & & & & & & & & & & & \\
\hline
Normalization & 1.019  & 1.057  & 0.965  &  1.004  &  0.946  &  1.030   & 0.956 &  1.051 & 1.004  & 0.0432 & 1.5\%  \\
 Slope & -1.937  & -2.208 &  -1.692  & -1.942  & -1.629  &  -2.027 &  -1.714  & -2.100 & -1.906  & 0.209  &  7.4\%  \hskip 0.6in  6.0\%  \\
N (139,798) &   20408    &   11973   &    21697  &     15821 &      15903    &   21615  &     11891  &     20490 & 17475 & 4129 &  \\
\hline
\end{tabular}
\end{center}
\label{cv_table}
\end{table}


\section{Epilog: Summary and Suggested Future Analysis}
\label{epilog}

The unadorned 3D Fourier transform of coordinates 
from a redshift survey 
can be used to characterize the spatial distribution
of galaxies, 
as demonstrated here 
with the volume limited sample defined in Papers I and II.
A simple Fourier sum over the 
galaxy positions
compares well with the transform 
of the same points 
binned in small 3D voxels.  
The direct sum has no 
resolution limit other than 
that inherent to the data
or due to computational limitations.
We display 3D Fourier power spectra,
as well as projections radially 
and in three orthogonal 
coordinate directions; projection in 
arbitrary directions 
could provide a straightforward
way to study isotropy.

However the emphasis here is on 
Fourier phase information,
of interest for example 
in the context of Gaussianity measures.
The phase spectrum 
has much to recommend it over the
more commonly used multi-point statistics 
and related methods.
We display maps 
(projected to 2D) of variance, 
kurtosis and entropy 
of nearest-mode phase differences 
to quantify distributional nonuniformity.
Such analysis of the SDSS data, 
taking into consideration 
simulated control samples and the
MS simulation,
has not provided any convincing 
evidence for non-uniformity in the distribution of phases.
This result is somewhat surprising 
since structure on any scale must generate
local Fourier non-uniformities and  
render the distribution of density values 
(in spatial voxels) non-normal 
\citep[e.g.][]{schaap}.
Furthermore even 
perfectly Gaussian initial density perturbations
should suffer evolutionary modifications
leading to nonGaussianity in the current distribution.
On the other hand, 
offsetting these effects  
are data-analytic issues 
such as the rather conservative measures
(e.g. phase differencing) we have been driven to,
the ill-defined nature of the target signal,
and the plethora of possibly relevant  
analysis methods -- only a tiny 
fraction of which has been explored here or in 
previous research by others.
In addition the normalization principle 
underlying the central limit theorem
is at work in samples of any size.

However we expect that improved data -- 
more recent SDSS data releases,
deeper selections of other existing surveys, 
new larger ones, 
compendia of several redshift surveys
covering similar redshift ranges, etc. -- 
and guidance from theory and simulations 
will elucidate these issues,
perhaps using phase analysis techniques 
augmented with the following 
three promising new approaches:


\subsection{Optimal Phase Bins via Bayesian Blocks}
\label{bb}

The first problem is finding a principle 
for defining sets of phases to analyze.
In an exploratory data analysis setting -- 
i.e. absent guidance from theoretical predictions -- 
one should, in principle at least, 
consider the set of all possible subsets.
A way to address the exponentially large 
size of such a collection 
is to use the Bayesian Block algorithm
\citep{jackson2005,scargle_vi} 
in its higher dimensional mode
\citep{jackson2010} 
to optimally partition the phases.
This  $O(N^{2})$ algorithm yields 
the optimal among the $2^{N}$ possible binnings,
where here 
the \emph{block cost function} to be optimized 
would be some NG metric for the data within
each block, for example kurtosis.
This leads to ...

\subsection{Independent Component Analysis}
\label{ica}

... the second problem: the choice of NG metric.
The close connection between independence
and nonGaussianity (cf. Section \ref{phase_spectrum}) 
suggests that \emph{independent component analysis} 
(ICA) will be useful.
The wide-ranging but 
comprehensive monograph by 
\cite{hyvarinen_book}  elucidates all of the 
NG issues discussed here and then some.
By elaborating its slogan
``nonGaussian is Independent''
this monograph
provides a unified picture of 
many inter-related properties of 
statistical processes -- including dependence, 
correlation, 
Gaussianity, non-linear correlation,
kurtosis, cumulants, and sparseness.
This book and the update 
\citep{hyvarinen} detail 
related algorithmic approaches
such as sparse coding, projection pursuit, 
principal component
and independent component analysis.
The existence of practical, fast ICA
algorithms\footnote{See https://research.ics.aalto.fi/ica/fastica/}
should facilitate application of these ideas 
to statistical cosmology.
We are thus led to ... 

\subsection{Large-Scale Inference: Higher Criticism and False Discovery Rate Control}
\label{lsi}
... address the last step, inference.\footnote{``Very roughly speaking,
algorithms are what statisticians do, while
inference says why they do them.'' \citep{efron_hastie}}
Recent advances in statistics have opened up 
a number of opportunities for future analysis 
of cosmological data.
What Brad Efron calls
``scientific mass production, in which new 
technologies ... allow a single team 
of scientists to produce [very large ] data sets ... "
has given birth to the field 
\emph{Large-Scale Inference} (LSI).
Two monographs \citep{efron_lsi,efron_hastie} review the 
statistical science underlying this discipline, 
its historical development, 
and its role in \emph{big data} contexts.
Two LSI techniques extremely popular in applied statistics, 
False Discovery Rate Control (FDRC) 
and Higher Criticism (HC),
address problems potentially of great importance
for large-scale astronomical data analysis.
In the generic setting, 
termed \emph{large-scale hypothesis testing},
one is faced with a large number of data elements, 
each providing evidence for or against a hypothesis 
(or possibly a different hypothesis for each datum).
The analysis techniques, much like
the trials factor,  focus on 
integrative issues such as assessing the probability
of making even one false rejection of a hypotheses 
in simultaneous analysis of $N$ hypothesis tests --
especially in cases where the signal is expected 
to be weak (individual ones  
may not be detectable on their own)
or rare (occurring in $\ll N$ of the cases). 

Higher Criticism, 
perhaps more informatively termed 
\emph{second-level significance testing},
was introduced by  \cite{donoho_jin_0},
following John Tukey's parable of the young psychologist,
further developed in 
\citep[e.g.][]{donoho_jin_0,donoho_jin_1,donoho_jin_2,walther2013}
and applied in cosmology
\citep[e.g.][]{cayon_jin,jin2005}.
The HC statistic may provide rigorous 
significance analysis of the multiple tests
which are implicit in phase maps -- effectively providing
a statistical \emph{trials factor} correction.
HC applied to numerous beams within the 
phase-cube could lead to HC 
analysis of the HC statistic itself -- \emph{third-level  significance testing} 
or \emph{even higher criticism}.
With somewhat different goals 
the FDR formalism, by controlling
the relative proportion of false discoveries,
can lead to more discoveries -- useful 
especially when follow-up of putative discoveries is 
practical.


\acknowledgements

We are grateful to the NASA-Ames Director's Discretionary Fund
and to Joe Bredekamp and the NASA Applied Information
Systems Research Program for support and encouragement.
Thanks goes to Ani Thakar and Maria Nieto-Santisteban for
their help with our many SDSS casjobs queries. Michael Blanton's
help with using his 
SDSS NYU--VAGC catalog is also very much appreciated.
We are also grateful to 
Chris Henze, 
Roger Blandford, 
Elliott Bloom,
Andrew MacFadyen, 
Jay Norris, 
Pratyush Pranav, 
Aaron Roodman, 
Alex Silbergleit, 
Luis Teodoro,
Bob Wagoner,
Emannuel Candes,
Peter Coles,
and 
Guenther Walther
for various helpful suggestions.
We especially thank the anonymous referee 
for comments that much improved this paper.

Funding for the SDSS has been provided by
the Alfred P. Sloan Foundation, the Participating Institutions, the National
Aeronautics and Space Administration, the National Science Foundation,
the U.S. Department of Energy, the Japanese Monbukagakusho, and the Max
Planck Society. The SDSS Web site is http://www.sdss.org/.

The SDSS is managed by the Astrophysical Research Consortium for
the Participating Institutions. The Participating Institutions are The
University of Chicago, Fermilab, the Institute for Advanced Study, the
Japan Participation Group, The Johns Hopkins University, Los Alamos National
Laboratory, the Max-Planck-Institute for Astronomy, the
Max-Planck-Institute for Astrophysics, New Mexico State University,
University of Pittsburgh, Princeton University, the United States Naval
Observatory, and the University of Washington.

This research has made use of NASA's Astrophysics Data System Bibliographic
Services.
\clearpage


\appendix

\section{Checking the Formalism
using the Inverse Fourier Transform}
\label{appendix_a}

It is useful to check how well our 
Fourier transform estimates 
capture the information in the galaxy coordinate data.
The discrete Fourier transform
of evenly spaced voxels is exactly invertible 
and therefore lossless, and so is the direct 
transform  in Eqs. (\ref{ft_a}) and (\ref{ft_b})
in the limit of an infinite number of frequencies.
Nevertheless it is of some interest to 
see how this limit is approached
by comparing its inverse transform 
against the raw data.
For this limited purpose
a rough visual check suffices, 
since a precise goodness-of-fit metric,
involving comparison
of an effectively continuous representation 
with point data, is difficult.

In the direct transform there is 
no binning of galaxy positions,
so if the Fourier transform were to be 
evaluated at an infinite number of 
spatial frequencies the inverse 
transform would exactly reproduce the
data. That is the function 
\begin{equation}
F_x(\bf{x} ) = 
\int F( {\bf k} ) e^{ -i \bf{k} \cdot \bf{x} } d\bf{k}
\end{equation}
\noindent 
would vanish except for unit delta functions
at each of the galaxy positions.
Normalization is not important here, so the
factor ${1 \over (2 \pi )^{3/2} }$ sometimes
written in front of the right-hand side of 
this equation is omitted.
Inserting Eq. (\ref{ft_a}) into this expression we have
\begin{equation}
F_x(\bf{x} ) = 
\sum_{n=1}^{N} \ \int  
e^{ -i \bf{k} \cdot ( \bf{x}  - \bf{x_n} ) } d\bf{k}
\end{equation}
\noindent 
It is well known that 
this integral is equivalent to 
a  $\delta-$function at $x_{n}$,
yielding the desired result.

\begin{figure}[htb]
\includegraphics[scale=1]{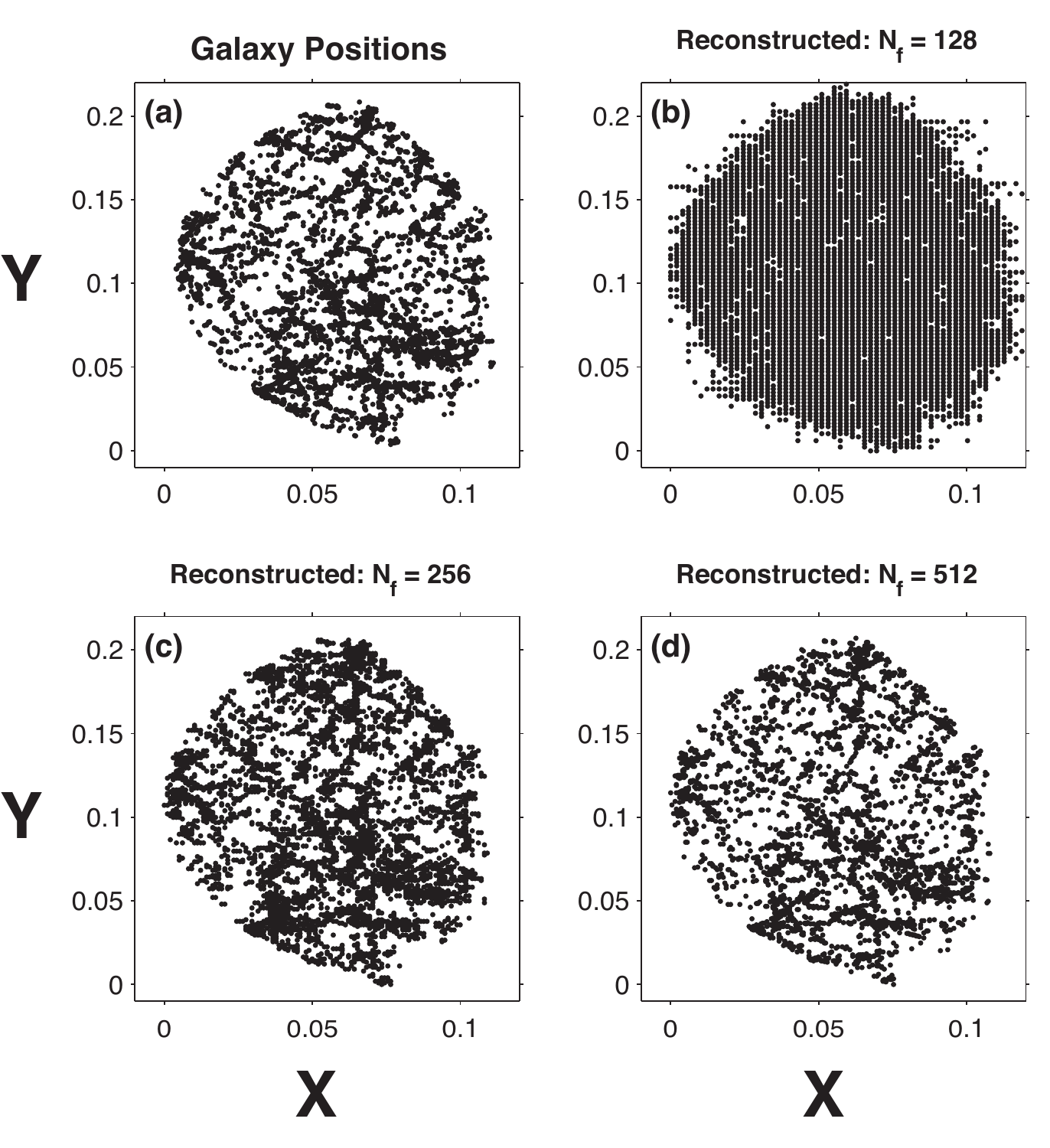}
\caption{Comparison of x-y projections of 
thin (12.5 Mpc.) z-slices  for
(a) the galaxy data;
and the corresponding reconstruction 
with the direct Fourier Transform in
Eq. (\ref{inverse_transform_abs}) using 
(b) 128 frequencies; (c) 256 frequencies,
and (d) 512 frequencies.
Coordinates are in redshift units (rsu).
The effective resolutions of the reconstructions
are 16.9, 8.5 and 4.2 Mpc, respectively.
Plots of other projections are very similar. }
\label{fig_appendix}
\end{figure}

It is useful to 
investigate how well these exact 
results apply to necessarily finite numerical computations. To do so we evaluate the 
expression
\begin{equation}
 f(x,y,z) =  \sum_{k_x,k_y,k_z} F(k_x, k_y, k_z )  
 e^{ -i ( k_x x + k_y y + k_z z ) }  
\label{inverse_transform_abs}
\end{equation}
\noindent 
appropriately normalized,
against the raw data 
in Figure \ref{fig_appendix}.
The total number of spatial frequencies 
increases as the third power of 
the number of frequencies in each dimension, 
but it is nevertheless feasible to 
use a frequency grid that well resolves
the relevant spatial structure in all three directions.
The spatial frequencies were taken to be the usual 
integer multiples of the fundamental frequency
of ${1 \over 1082}  Mpc^{-1}$. This denominator is 
approximately twice the maximum of the x, y and z
ranges of the data, to eliminate wraparound.

Since the forward transform can be evaluated
at any set of spatial frequencies, it is expedient
to use an FFT algorithm\footnote{Our 
expression for the forward transform
does not automatically impose the 
complex conjugate symmetry necessary for
the inverse transform computed in this way
to be real.  To deal with this problem
we simply evaluate the forward transform
at an odd number of points: one corresponding
to zero frequency, $(N-1)/2$ at positive
frequencies, and the remaining $(N-1)/2$ at
the corresponding negative frequencies.
This symmetry yields a positive result.
Accordingly values for the number of frequencies
are written in the form $N+1$ throughout,
where $N$ is even.}
to evaluate the expression above.
We reconstructed 3D data points 
at every location
in the $ N_{k} \times N_{k} \times N_{k} $ array 
($N_{k}^{3}$ voxels) 
where the value of $f(x,y,z)$ in equation (\ref{inverse_transform_abs}) 
exceeds a threshold.
This threshold value was 
chosen to yield the same number of points 
(139,798) as in the raw data.
The figure shows xy-projections of
the points contained in a  
12.5 Mpc thick slice in the z-coordinate,
isolating roughly 5,000 points in all three panels.
The limited frequency range 
dictates that the reproductions are 
smoothed representations of the galaxy data.
The sequence in this figure 
demonstrates that increasing the number of 
spatial frequencies reproduces the 
discrete raw data with improved accuracy.
Note that panel (c) with $N_{k} = 256+1$ seems
to have more points than (d) for $N_{k} = 512+1$;
in fact both have the same number,
those in (d) more closely following the narrow
filaments and other structures (with a consequent
increase in overplotting of points) and therefore 
more faithfully reproducing the data. 
The key point is 
that information about 
the discrete structure at a broad range of scales,
limited only by the resolution of the computation,
is contained in
the Fourier transform in eq. (\ref{ft_b}).

\clearpage

\begin{figure}[htb]
\includegraphics[scale=1]{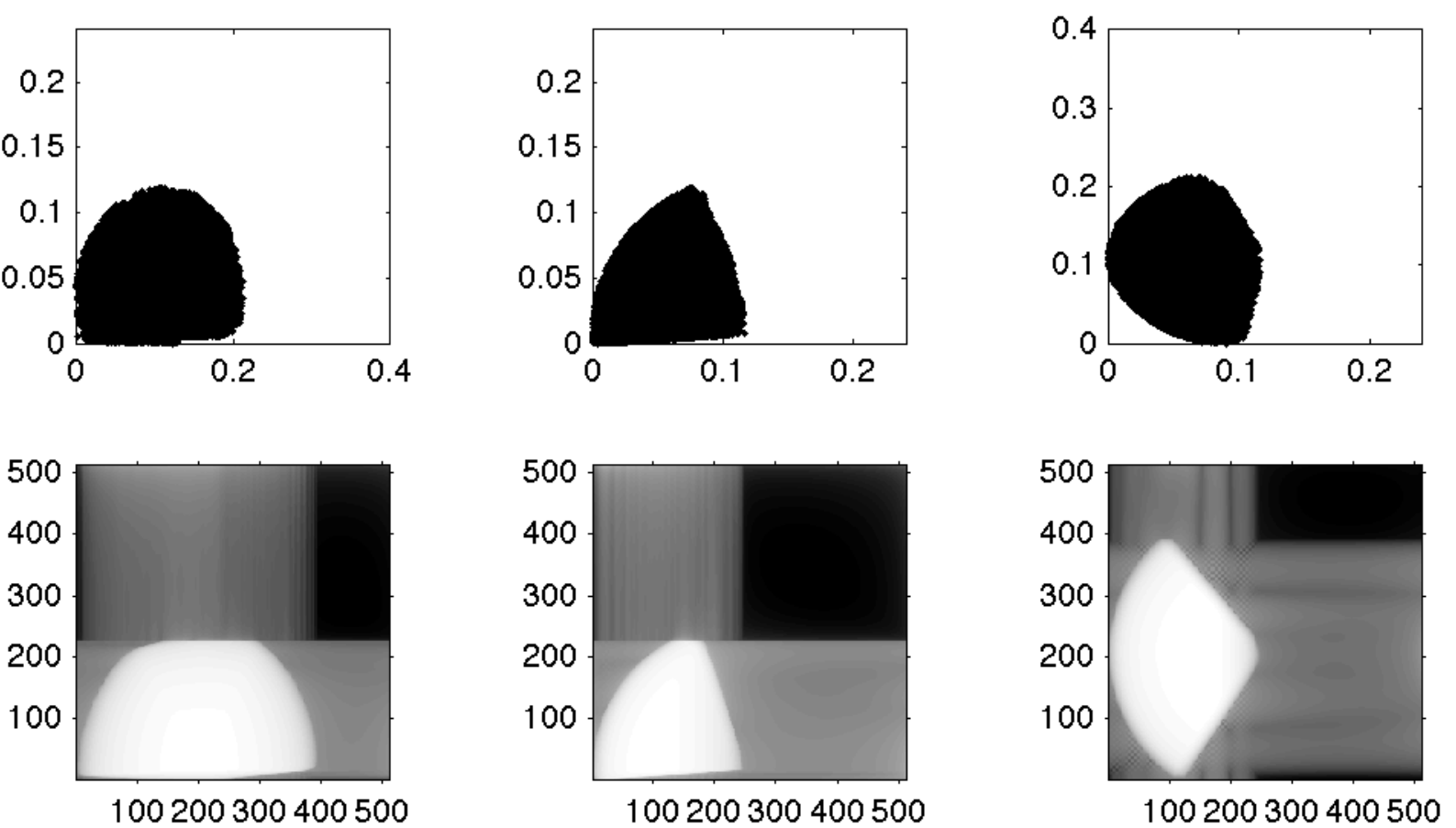}
\caption{Reconstruction of the selection function.
Top row: projections of the raw galaxy coordinates
along the x- , y- , and z-axes respectively
(empty white spaces indicate the extended ranges 
employed in the Fourier transforms)
for comparison with the corresponding
reconstructions via the 
inverse Fourier transform 
in the bottom row.
These are 2D projections of thin slices
oriented as indicated by the axis labels,
with the gray-scale representing the reconstructed 
density.
Here the grid size is .001 redshift units,
for which the relative volume error 
in the cuboid representation is
about one part in ten thousand.} 
\label{window_ifft}
\end{figure}

In the same way
the projected density plots in Figure \ref{window_ifft} 
demonstrate that the inverse transform of the
selection function Fourier transform 
is essentially a uniform solid 
corresponding to the observational data space.
This procedure accounts for incorporating
only galaxies inside the window, 
but of course does not in any way replace or estimate data outside of the window.


\section{3D Fourier Transforms: MatLab Code}
\label{appendix_b}

MatLab code 
(\verb+doi:10.5281/zenodo.432820+) 
computes the Fourier transform of the 
galaxy coordinates, the corresponding window function
and its deconvolution.
The former is a direct evaluation of Equation \ref{ft_b};
the latter is based on a refined partition of the actual
data space -- here taken to be the 
convex hull of the galaxy positions -- into cuboids
with $x$ and $y$ coordinates in an evenly spaced
rectangular grid, as described in Section \ref{3D_window}.
More details are 
available in the ReadMe file and commented
MatLab scripts at 
\verb+https://doi.org/10.5281/zenodo.432820+

\end{document}